# KURVS: chemical properties from multiple strong line calibrations for star-forming galaxies at $z \sim 1.5$

Zefeng Li ,[1]★ Ugnė Dudzevičiūtė ,[2,3,4] Annagrazia Puglisi ,[5] Steven Gillman ,[3,4]
A. Mark Swinbank ,[1] Luca Cortese ,[6] Ian Smail ,[1] Karl Glazebrook ,[7] Anna F. McLeod ,[1]
Dominic J. Taylor ,[1] Roland Bacon ,[8] Christopher Harrison ,[9] Edo Ibar ,[10,11] Juan Molina ,[10,11]
Danail Obreschkow [6] and Tom Theuns [12]

[1]*Centre for Extragalactic Astronomy, Department of Physics, Durham University, South Road, Durham DH1 3LE, UK*
[2]*Max – Planck – Institut für Astronomie, Königstuhl 17, D-69117 Heidelberg, Germany*
[3]*Cosmic Dawn Center (DAWN), Copenhagen, Denmark*
[4]*DTU-Space, Technical University of Denmark, Elektrovej 327, DK-2800 Kgs. Lyngby, Denmark*
[5]*School of Physics and Astronomy, University of Southampton, Highfield SO17 1BJ, UK*
[6]*International Centre for Radio Astronomy Research, University of Western Australia, 35 Stirling Highway, Crawley, WA 6009, Australia*
[7]*Centre for Astrophysics and Supercomputing, Swinburne University of Technology, PO Box 218, Hawthorn, VIC 3122, Australia*
[8]*Univ Lyon, Univ Lyon1, ENS de Lyon, CNRS, Centre de Recherche Astrophysique de Lyon UMR5574, 9 avenue Charles André, F-69230 Saint – Genis – Laval, France*
[9]*School of Mathematics, Statistics and Physics, Newcastle University, Newcastle upon Tyne NE1 7RU, UK*
[10]*Instituto de Física y Astronomía, Universidad de Valparaíso, Avda. Gran Bretaña 1111, Valparaíso, Chile*
[11]*Millennium Nucleus for Galaxies (MINGAL), Avda. Gran Bretaña 1111, Valparaíso, Chile*
[12]*Institute for Computational Cosmology, Department of Physics, Durham University, South Road, Durham DH1 3LE, UK*

## ABSTRACT

Gas-phase oxygen abundance (metallicity) properties can be constrained through emission line analyses, and are of great importance to investigate galaxy evolution histories. We present an analysis of the integrated and spatially resolved rest-frame optical emission line properties of the ionized gas in 43 star-forming galaxies at $z \sim 1.5$ in the *K*-band Multi Object Spectrograph Ultra-deep Rotational Velocity Survey. Using the [N II]$\lambda6584$/H$\alpha$ (N$_2$), ([O II]$\lambda\lambda3727$, 9+[O III]$\lambda\lambda4959$, 5007)/H$\beta$ (R23), and for the first time [N II]$\lambda6584$/[O II]$\lambda\lambda3727$, 9 (N$_2$O$_2$) indicators at this redshift, we measure the gas-phase metallicities and their radial gradients. On $\sim$ 4-kpc scales metallicity gradients measured from N$_2$O$_2$ and those measured from N$_2$ are in good agreement when considering the spatial distributions of dust in each galaxy, as parametrized by dust attenuation radial gradients. We report a nearly flat metallicity gradient distribution typically at $z \sim 1.5$, with the 50th, 16th, and 84th percentiles at 0.01, $-0.03$, and 0.05 dex kpc$^{-1}$, respectively. The findings agree well with previous observational studies and simulations at this epoch. We ascribe the observed negative metallicity gradients to a natural result from self-regulating systems, and the positive ones to potential galactic fountains and higher merger rates.

## 1 INTRODUCTION

The measurement of gas-phase chemical abundance in the interstellar medium (ISM) in galaxies reflects the complex interplay between gas accretion, star formation, and outflows (the baryonic cycle; e.g. B. M. Tinsley 1980). The abundance of oxygen, the most abundant heavy element (except hydrogen and helium) in the Universe, is defined as a common proxy for total gas phase metallicity. The average metallicity of a galaxy is closely correlated with stellar mass ($M_*$), a scaling relation commonly referred as the stellar mass–metallicity relation (MZR; C. A. Tremonti et al. 2004). The MZR has been extended to lower-stellar-mass galaxies, and is now established from

$10^7$ to $10^{12}$ M$_\odot$, with a steep slope at low masses that flattens at higher masses (D. K. Erb et al. 2006; H. Lee et al. 2006; F. Mannucci et al. 2010; E. Pérez-Montero 2014; J. H. Lian et al. 2015). Considering that metals are products from star formation, the MZR is considered as a sequence of galaxy evolutionary stages (R. Maiolino et al. 2008; H. J. Zahid, L. J. Kewley & F. Bresolin 2011). The MZR has been also confirmed from the local Universe to $z \sim 8$ (e.g. E. Wuyts et al. 2012; A. Henry et al. 2013; H. J. Zahid et al. 2014; K. Yabe et al. 2015; D. Kashino et al. 2017; T. Jones et al. 2020; M. W. Topping et al. 2021), and to $z \sim 10$ after the launch of the *James Webb Space Telescope* (*JWST*; e.g. K. Nakajima et al. 2023; I. Chemerynska et al. 2024; M. Curti et al. 2024).

Beyond the integrated metallicities, distributions of metals within galaxies provide important information of star formation histories. The most well-studied distribution is the radial metallicity gradi-ent. Since the age of individual nebula observations and long-slit spectroscopy, pioneering studies (e.g. L. Searle 1971; G. A. Shields 1974; B. E. J. Pagel, M. G. Edmunds & G. Smith 1980) have reported the presence of negative metallicity gradients in spiral galaxies by measuring strong emission lines from H II regions. After the application of long slit, seminal surveys (M. B. Vila-Costas & M. G. Edmunds 1992; D. Zaritsky, J. Kennicutt & J. P. Huchra 1994; R. B. C. Henry & G. Worthey 1999) then confirmed the negative metallicity gradients in spiral galaxies typically at −0.05 to −0.1 dex kpc⁻¹, also using empirical strong line indicators. More recent surveys utilizing integral field units (IFUs) on 4–10 m class telescopes have obtained resolved optical spectroscopy for thousands of nearby galaxies (e.g. S. M. Croom et al. 2012; S. F. Sánchez et al. 2012; K. Bundy et al. 2015; S. Erroz-Ferrer et al. 2019; C. López-Cobá et al. 2020; E. Emsellem et al. 2022). The IFU data sets provide 2D spatially resolved metallicity maps revealing radial variations in chemical abundances within galaxies (e.g. S. F. Sánchez et al. 2014; I. T. Ho et al. 2015; F. Belfiore et al. 2017; L. Sánchez-Menguiano et al. 2017; H. Poetrodjojo et al. 2018; K. Kreckel et al. 2019). The radial gradients from spatially resolved spectroscopy trace the chemical enrichment and star formation histories in discs, providing key constraints on galaxy formation models, one of which is the confirmation of the inside–out formation scenario (C. Chiappini, F. Matteucci & R. Gratton 1997; C. Chiappini, F. Matteucci & D. Romano 2001), with rapid enrichment and earlier stellar buildup in galaxy centres and longer accretion time-scales at larger radii. Following this scenario, many analytical models build upon the 'gas regulator' framework by applying it to concentric radial zones in a galaxy. In these multizone models, key parameters such as gas infall time-scales and star formation efficiencies are specified as a function of galactocentric radius (e.g. M. Mollá, F. Ferrini & A. I. Díaz 1997; T. Naab & J. P. Ostriker 2006; A. Mott, E. Spitoni & F. Matteucci 2013; E. Spitoni, F. Matteucci & M. M. Marcon-Uchida 2013; E. Spitoni et al. 2015; P. Sharda et al. 2021).

---

★ E-mail: zefeng.li@durham.ac.uk

Metallicity gradient studies have also been extended to $z \sim 4$, thanks to high spatial resolution instruments and gravity lens amplification, though the cosmic dimming $\propto (1 + z)^4$ is a great challenge for any study beyond the local Universe. A number of studies (e.g. J. Queyrel et al. 2012; A. M. Swinbank et al. 2012; P. Troncoso et al. 2014; N. Leethochawalit et al. 2016; E. Wuyts et al. 2016; D. Carton et al. 2017; J. Molina et al. 2017; X. Wang et al. 2017, 2019; M. Curti et al. 2020; S. Gillman et al. 2021) have reported metallicity gradients at $z < 4$. These studies have demonstrated a variety of metallicity gradients at $z \gtrsim 1$, though the majority of them reported nearly flat ones. However, the metallicity indicators are limited in [N II$\lambda$6584]/H$\alpha$, accessible at high redshifts and insensitive to dust correction, but sensitive to other ionized gas properties (e.g. ionization parameter, a dimensionless quantity to describe the ratio of ionizing photon density to gas density). Various metallicity calibrations are needed to confirm the robustness of the measured metallicity gradients against other gas properties.

Recently, the *K*-band Multi Object Spectrograph (KMOS) Ultra-deep Rotational Velocity Survey (KURVS; A. Puglisi et al. 2023) aims to make full use of the KMOS on the Very Large Telescope (VLT) to measure the rotation curves of ionized gas in a representative sample of 43 main-sequence star-forming galaxies at $z \sim 1.5$. This represents one of the largest sample size to date at this redshift, allowing a comprehensive analysis of the connection between galactic chemical evolution and the kinematics of ionized gas. By including multiband observations in the *IZ*, *YJ*, and *H* bands that enables the simultaneous coverage of many rest-frame optical emission lines, in this paper we derive both integrated and spatially resolved metallicities using different strong-line diagnostics for a large sample at $z \sim 1.5$. The utilization of various metallicity indicators beyond [N II$\lambda$6584]/H$\alpha$ offers more reliable measurements of chemical abundances.

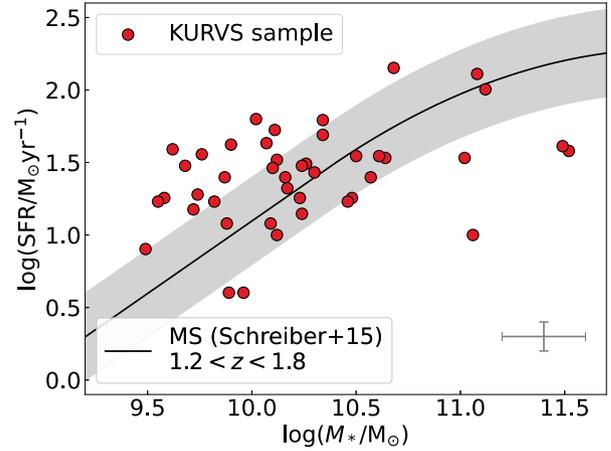

**Figure 1.** Star formation rate versus stellar mass for the KURVS sample of 43 galaxies shown as circles. The solid line and the band show the best fit and main sequence range at $z \in (1.2, 1.8)$ from C. Schreiber et al. (2015), respectively. The typical error bars of the two physical quantities are shown in the cross marker in the lower right corner.

The outline of this paper is as follows: In Section 2, we introduce the observation, sample selection, and data reduction. We report the analysis and discuss the results of integrated and spatially resolved properties of the sample in Sections 3 and 4, respectively. We draw the conclusion in Section 5. Throughout this paper we adopt a cosmology defined by $H_0 = 67.3 \text{ km s}^{-1} \text{ Mpc}^{-1}$ and $\Omega_m = 0.315$ (Planck Collaboration XVI 2014).

## 2 SAMPLES, OBSERVATIONS & DATA REDUCTION

The 43 targeted KURVS galaxies lie in the Cosmic Assembly Near-infrared Deep Extragalactic Legacy Survey (N. A. Grogin et al. 2011) fields. They are distributed in two separated fields: the Chandra Deep Field South (CDFS) field and the Cosmological Evolution Survey (COSMOS) field. The 20 galaxies in the CDFS field and the 14 galaxies in the COSMOS field are selected from the KMOS Galaxy Evolution Survey (S. Gillman et al. 2020; A. L. Tiley et al. 2021). Two galaxies in the CDFS field and seven galaxies in the COSMOS field are selected from the KMOS³D survey (E. Wisnioski et al. 2019). In each field 22 targets along with one reference star (to monitor point spread function, PSF) are simultaneously observed making full use of the KMOS's 24 IFUs. We report the integrated properties of our sample of the 43 galaxies in Table A1 (with a horizontal line separating the galaxies in the CDFS and COSMOS fields).

The KURVS galaxies have $z_{spec}$ ranging from 1.23 to 1.71 with a median value of $z_{spec} = 1.50$. The stellar masses and star formation rates are derived from *JWST* and the Hubble Space Telescope (*HST*) photometry energy distribution (SED) fitting from the CDFS (Y. Guo et al. 2013) and COSMOS (A. Muzzin et al. 2013) photometric catalogues using MAGPHYS (E. da Cunha, S. Charlot & D. Elbaz 2008; E. da Cunha et al. 2015). The median stellar mass of the KURVS samples is $\log(M_*/M_\odot) = 10.2$, and the median dust-corrected star formation rate is 22 M$_\odot$ yr⁻¹ (Fig. 1). The morphological parameters (position angles and inclination angles) are derived from *JWST* and *HST* broad-band images (Dudzevičiūtė et al. in preparation), assuming a thick disc model with $q_0 = 0.2$.[1]

The KURVS galaxies were observed in the $H$, $YJ$, and $IZ$ bands between October 2018 and December 2023. The $H$-band data were obtained from 2019 to 2021, and the total exposure time is $\sim 80\,\mathrm{h}$ each, while the $YJ$- and $IZ$-band data were obtained from 2021 to 2023, and the total exposure time is $\sim 20\,\mathrm{h}$ each. At $z \sim 1.5$ the $H$ band (1.46–1.85 μm) covers the emission lines of Hα, [N II]λλ6548, 84, and [S II]λλ6717, 31, the $YJ$ band (1.02–1.34 μm) covers Hβ, [O III]λλ4959, 5007, and the $IZ$ band (0.78–1.08 μm) covers [O II]λλ3727, 9. The typical seeing during the observing periods is $\sim 0.6$.

Data cubes of KMOS targets have been reconstructed using the European Southern Observatory Recipe Execution Tool ESOREX pipeline. The pipeline conducts standard dark, flat, and wavelength calibrations. Stacks are made from the median of the individual observations with spatial offsets measured from the reference stars.

## 3 INTEGRATED PROPERTIES

In this section, we first focus on the galaxy integrated emission line fluxes. We integrate the individual spectra within an elliptical aperture with 1.5 times the half-light radius ($R_e$, median value for the KURVS sample is 3.7 kpc; A. Puglisi et al. 2023).

When integrating the individual spectra of the spaxels within the aperture, we remove the continua and adopt the correction for galaxy rotational velocities to enhance the gain of signal-to-noise ratios (S/Ns) of line fluxes. We use the velocity maps from Dudzevičiūtė et al. (in preparation) at the same spatial sampling (0.1 arcsec). If one spaxel has no robustly measured velocity, we assign that of the closest available spaxel to maximize the number of the integrated spaxels. We apply the fitting scheme to the velocity-corrected integrated spectra and we show the spectra and fitting results of all the KURVS galaxies in Appendix A.

We model the emission line properties using a $\chi^2$ minimization fitting of both Gaussian components and flat continua. At each band the emission lines share the same line width and fixed relative positions. We consider sky emissions by including the reciprocal of the KMOS sky spectrum as weighting fraction. The line flux ratio between [N II]λ6584 and [N II]λ6548 is fixed at 3 and so is that between [O III]λ5007 and [O III]λ4959 (J. M. Vilchez & C. Esteban 1996). We take the intrinsic instrument spectral dispersion into account. The median full width at half-maximum (FWHM) of the emission lines are equivalent to $\sim 150\,\mathrm{km\,s^{-1}}$ in our sample. Our fitting scheme will be applied to both integrated spectra in Section 3 and spatially resolved data cube in Section 4. We also apply a statistical fitting that sums all the signals over a bandwidth of the FWHM, regardless of the emission profile. The S/Ns from the statistical fitting is typically lower than those from the least-$\chi^2$ fitting, due to the latter will not be affected by any absorption features. We find that a threshold S/N of 3 from the statistical fitting corresponds to a typical threshold S/N of 4 from the least-$\chi^2$ fitting, such that we adopt S/N > 4 as the minimum threshold throughout the paper.

In the KURVS we achieve a median S/N at 54 for Hα, 4.0 for Hβ, 14 for [N II]λ6584, 7.0 for [O III]λ5007, and 4.3 for [O III]λλ3727, 9, since the exposure time at the $H$ band is much longer than that at the other bands. A. Puglisi (2023) report two identified active galactic nucleus (AGNs) in the KURVS sample (KURVS_11 and KURVS_12), from their infrared colours. In this analysis, KURVS_11 does not have reliable detection at the $YJ$ and $H$ bands, and KURVS_12 has very broad components in the Hα emission (see Fig. A2). We emphasize additional caution paid to these two targets in the following chemical abundance analyses.

### 3.1 BPT diagram

Since our data consist of two groups in terms of observational depths – total exposure of $\sim 80\,\mathrm{h}$ in the $H$ band and $\sim 20\,\mathrm{h}$ in the $YJ$- and $IZ$-bands, as a first calibration of data quality we plot line ratio [O III]λ5007/Hβ as a function of [N II]λ6584/Hα, i.e. the Baldwin–Phillips–Terlevich diagram (BPT diagram; J. A. Baldwin, M. M. Phillips & R. Terlevich 1981), and of stellar masses in Fig. 2. For [O III]λ5007, Hβ, [N II]λ6584, and Hα, we require a minimum S/N of 4 for all lines. Since Hβ is the faintest line among them, we also include galaxies with S/N below this threshold, representing them with small light grey circles and upwards arrows. For these cases, we adopt 4σ upper limits and plot them in Fig. 2.

For comparison with other studies, we extracted a sample of local galaxies with estimated line fluxes and stellar masses from the Sloan Digital Sky Survey (SDSS, the MPA-JHU catalogue; G. Kauffmann et al. 2003a; J. Brinchmann et al. 2004). The location of KURVS galaxies on the BPT diagram is above the majority of that of the SDSS galaxies, and in a good agreement with a series of studies at $z$ from $\sim 1.5$ to $\sim 3$ (C. C. Steidel et al. 2014; H. J. Zahid et al. 2014; A. E. Shapley et al. 2015, 2024; A. L. Strom et al. 2017; D. Kashino et al. 2019). In the right panel of Fig. 2, we show the ratios of log([O III]λ5007/Hβ) as a function of stellar mass. We also plot the best-fitted linear in (A. L. Strom et al. 2017, $z \sim 2.3$) and (D. Kashino et al. 2019, $z \sim 2.3$), and from the SDSS local galaxy sample. This illustrates the similar result that the KURVS galaxies locate above the SDSS galaxies and the demarcation built in the local Universe (S. Juneau et al. 2014).

Due to the small number of data points (< 10) we do not report the best fits in this work. Qualitative comparisons suggest that our sample lies close to the best-fitting lines or curves at similar redshift, and above the majority of the sample in the local Universe, in terms of the BPT diagrams. The findings confirm the feasibility of combining the observations of different bands at various observational depths.

### 3.2 Oxygen abundances

In this subsection, we report the integrated gas-phase oxygen abundances for the KURVS sample. First, we discuss the dust correction, as some metallicity diagnostics rely on emission lines that are widely separated in wavelength, making them sensitive to dust reddening. We show the following analysis to demonstrate our choice of dust correction. In the gas phase, the dust dereddening correction can be measured using the intrinsic Balmer decrement Hα/Hβ = 2.86 (line flux ratio) assuming a case B recombination at electron temperature $T_e = 10\,000\,\mathrm{K}$ and density $n_e = 100\,\mathrm{cm^{-3}}$ (D. E. Osterbrock 1989). The gas-phase dust attenuation is provided as

$$A_{\mathrm{V,gas}} = R_V E(B - V)$$
$$= \frac{R_V}{0.4(K_{H\beta} - K_{H\alpha})} \times \log\left(\frac{H\alpha/H\beta}{2.86}\right), \tag{1}$$

where $K_{H\alpha} = 2.46$ and $K_{H\beta} = 3.82$ are adopted from the average dust attenuation curve at $z \sim 1.3$ (A. J. Battisti et al. 2022, $R_V = 3.22$). In the stellar phase we obtain the dust attenuation $A_{V,\text{star}}$ from MAGPHYS SED fitting.[2]

---

[1]The inclination angle $i$ satisfies $\cos^2 i = \frac{(b/a)^2 - q_0^2}{1 - q_0^2}$, where $b/a$ is the axial ratio in the image and $q_0 = 0.2$.

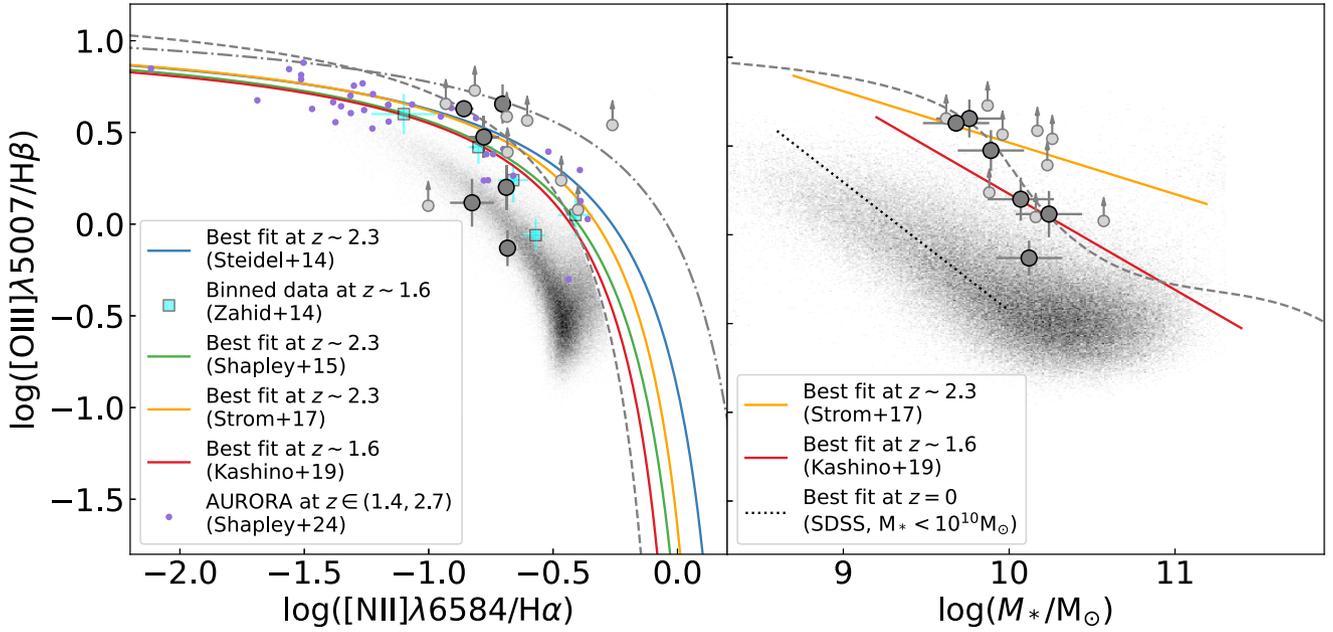

**Figure 2.** $\log([O\,\textsc{iii}]/H\beta)$ as a function of $\log([N\,\textsc{ii}]/H\alpha)$ (left, BPT diagram) and stellar mass (right) for all the KURVS galaxies with good (S/N > 4) detection of $[O\,\textsc{iii}]\lambda5007$, $H\beta$, $[N\,\textsc{ii}]\lambda6584$, and $H\alpha$. The large dark circles show the galaxies with S/Ns of all the four emission lines above 4, and the light small circles with upwards arrows show the lower limits of the $\log([O\,\textsc{iii}]/H\beta)$ line ratios by replacing the $H\beta$ line fluxes of which the S/Ns are below 4 by $4\times$ their noises. In the left panel, the blue, green, orange, and red curves show the best fits of C. C. Steidel et al. (2014), A. E. Shapley et al. (2015), A. L. Strom et al. (2017), and D. Kashino et al. (2019), respectively. The cyan squares show the mass-binned data of H. J. Zahid et al. (2014) and the purple dots show the AURORA survey from A. E. Shapley et al. (2024). The grey dashed and dotted curves show the demarcations of star-forming galaxies (G. Kauffmann et al. 2003b) and the theoretical maximum starburst galaxies (L. J. Kewley et al. 2001) in the local Universe. The background shows the probability density map for the SDSS galaxies. In the right panel grey dashed line shows the division between between star-forming/composite galaxies and AGNs at $z\sim0.1$ proposed by S. Juneau et al. (2014). The dotted line shows the best fit for SDSS low-mass galaxies. The best-fitting curve and line in the two panels agree well with the previous studies at $z\in(1.4,2.7)$. The two panels both show that the results of our sample are in good agreement with the previous studies at similar redshift ranges.

In this experiment, we only discuss the subset of six galaxies in our sample with S/Ns > 4 in both $H\alpha$ and $H\beta$ and $H\alpha/H\beta$ line flux ratios larger than 2.86. It is also possible that $H\alpha/H\beta$ line ratios are smaller than 2.86, which we do not take into account since the detailed physics are beyond the scope of this analysis. For the six galaxies we compute their dust dereddening based on equation (1), and the uncertainty is given by the same bootstrapping procedure described in Section 3.1. In Fig. 3, we show the $A_{V,\text{gas}}$ as a function of $A_{V,\text{star}}$ for the six individual galaxies where we have the two values simultaneously. In Fig. 3, we also show $A_{V,\text{gas}}$ of the stellar-mass-weighted stacked spectrum and $A_{V,\text{star}}$ taken from the averaged value of the galaxies, for the six galaxies (the dark blue square) and all the galaxies in our sample (the red square). S. Wuyts et al. (2013) report a conversion prescription from stellar to gas dust attenuation as $A_{V,\text{gas}} = A_{V,\text{star}}(1.9 - 0.15A_{V,\text{star}})$. It shows in Fig. 3 that the S. Wuyts et al. equation overestimates the gas dust attenuation from the stacked spectrum by a factor of $\sim1$ magnitude. The large scatter of our sample means that it is not trivial to directly infer $A_{V,\text{gas}}$ from $A_{V,\text{star}}$ for the individual galaxies. However, the averaged $A_{V,\text{gas}}$ derived from the stacked spectrum of all the galaxies agrees well with both the six galaxies with good detection in $H\beta$ emission lines and the previous studies on averaged $A_{V,\text{gas}}$ derived from Balmer decrement (A. Domínguez et al. 2013; A. J. Battisti et al. 2022; J. Matharu et al. 2023, adopting the same A. J. Battisti et al. attenuation curve). Thus, we adopt the Balmer-decrement-inferred gas-phase dust attenuation for the six galaxies, and the constant dust attenuation of $A_{V,\text{gas}} = 0.38 \pm 0.15$ from the stacked spectrum of all the galaxies for the rest in our sample.[3] We note that this uncertainty has been considered in the following metallicity estimates from line ratios.

Given the various strong emission lines accessible in KURVS, we estimate the metallicities using a variety of strong metallicity indicators. In this work, we mainly consider three strong line diagnostics: $[N\,\textsc{ii}]\lambda6584/H\alpha$ (N2; F. Bian et al. 2018, originally from M. Pettini & B. E. J. Pagel 2004; R. A. Marino et al. 2013), $([O\,\textsc{ii}]\lambda3727, 9+[O\,\textsc{iii}]\lambda\lambda4959, 5007)/H\beta$ (R23; R. Maiolino et al. 2008, originally from D. Zaritsky et al. 1994; H. A. Kobulnicky & L. J. Kewley 2004), and $[N\,\textsc{ii}]\lambda6584/[O\,\textsc{ii}]\lambda\lambda3727, 9$ (N2O2; L. J. Kewley et al. 2019, originally from L. J. Kewley & M. A. Dopita 2002). These indicators have been recently recalibrated by detection of auroral lines using *JWST* (e.g. R. L. Sanders et al. 2024) in studies (e.g. I. H. Laseter et al. 2024; E. Cataldi et al. 2025; D. Scholte et al. 2025) that focus on redshifts much higher ($2 < z < 9$) than this work. Among the metallicity indicators mentioned above, the N2 indicator is one of the most commonly used one especially at high-$z$ given the accessibility of wavelength coverage. However, the N$_2$ diagnostic is sensitive to ionization parameter. The R23 indicator does not rely on nitrogen emission lines but is also sensitive to ionization parameter and double-valued at $12 + \log(\text{O/H}) \sim 8.1$. N$_2$O$_2$[4] is one of the most reliable diagnostics as it is the least sensitive to ionization parameter [log($U$), the ratio of the local ionizing photon flux and the local hydrogen density] and marginal dependence on ISM pressure (L. J. Kewley et al. 2019). However, any indicators based on N$_2$O$_2$ actually measures the N/O ratio and then relying on the fact that at $12 + \log(\text{O/H}) \gtrsim 7.7$ N/O is positively correlated with O/H (at both low- and high-$z$; e.g. B. H. Andrews & P. Martini 2013; C. C. Steidel et al. 2014; C. Hayden-Pawson et al. 2022). Though N$_2$O$_2$ is reliable, it is still of great importance to compare the metallicities from N$_2$O$_2$ with those from the indicators that do not require nitrogen emission lines (e.g. R23). In Appendix B, we show the covariances to illustrate the discrepancy between any pair of metallicity indicators. In Fig. B1, the metallicities from N$_2$O$_2$ are in reasonable consistency with those form R23, though the sample is limited in fewer than five galaxies.

---

[2] *HST* F336W, F435W, F606W, F775W, F814W, F850LP, F105W, F125W, F140W, F160W, *Spitzer Space Telescope* 3.6 and 4.5 μm for KURVS_1 to KURVS_22 in the CDFS field; *HST* F606W, F814W, *JWST* F090W, F115W, F150W2, F200W, F277W, F356W, and F444W for KURVS_23 to KURVS_44 in the COSMOS field.

[3] The median and mean stellar mass of the six galaxies are both smaller than those of the whole sample by $\lesssim0.25$ dex, comparable to the 0.2 dex intrinsic measurement uncertainty, suggesting the stacked $A_{V,\text{gas}}$ representative in the following analysis.

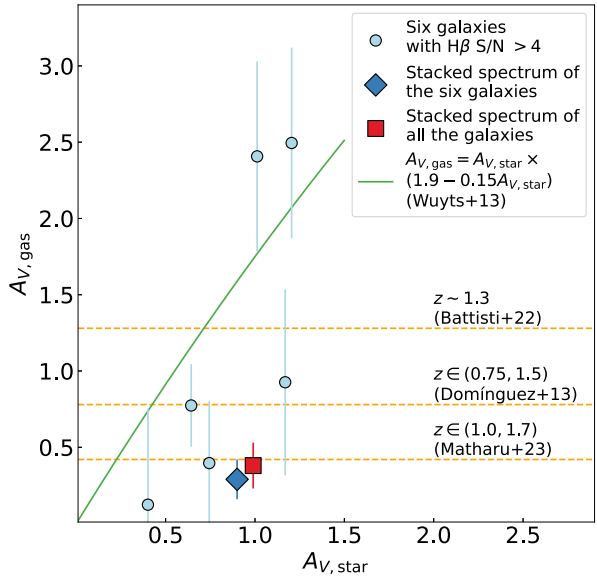

Figure 3. Gas-phase dust attenuation as a function of stellar-phase dust attenuation for the KURVS galaxies whose H$\alpha$/H$\beta$ line ratios are larger than 2.86. The circles show the galaxies with S/Ns of H$\beta$ line fluxes larger than 4. The diamond shows $A_{V,\text{gas}}$ derived from the stellar-mass-weighted stacked spectrum of the six galaxies versus the mean $A_{V,\text{star}}$ of the six galaxies. The square shows the same physical quantity as the diamond does, but for all the galaxies in our sample. The horizontal lines indicate the constant $A_{V,\text{gas}}$ in the literature (A. Domínguez et al. 2013; A. J. Battisti et al. 2022; J. Matharu et al. 2023) at similar redshifts and stellar mass bins centred on the median value ($\sim 10^{10}$ M$_\odot$) of our sample. The solid curve illustrates the best fit in S. Wuyts et al. (2013) from the extra dust correction for multiple star formation rate indicators.

We analyse the well-known gas-phase metallicity dependence on stellar mass for the KURVS galaxies using the three available metallicity indicators. In Fig. 4, we show the integrated metallicities as a function of stellar mass of all the three available indicators (individual galaxies appear in more than one panel if their metallicities could be calculated) and their best fits, compared with previous MZR studies (N$_2$ from R. Maiolino et al. 2008 converted to N$_2$ from F. Bian et al. 2018). We adopt the following equation as our MZR,

$$12 + \log(\text{O/H}) = Z_0 - \log\left[1 + \left(\frac{M_*}{M_0}\right)^{-\gamma}\right]. \tag{2}$$

In the equation $Z_0$ is the saturated metallicity at the massive end, $M_0$ is the turnover stellar mass above which the relation flattens and asymptotically approaches $Z_0$, and $\gamma$ is the slope at which the relation increases below $M_0$. Given the sample sizes, we only report the best-fitting parameters for the N$_2$ indicator as $Z_0 = 9.00 \pm 0.05$, $\log(M_0/\text{M}_\odot) = 10.87 \pm 0.30$, and $\gamma = 0.32 \pm 0.06$. The uncertainties are estimated using the bootstrapping procedure described in Section 3.1, setting the error of stellar mass as 0.2 dex (typical error in MAGPHYS and shown in Fig. 1).

In Fig. 4, our results from N$_2$ agree well with the previous work at the similar redshift range given the typical uncertainties. Our results from N$_2$ are flatter and higher than those in H. J. Zahid et al. (2014), and at the low stellar mass end our results are in better agreement with the results in S. Wuyts et al. (2013). For the R23 indicator our results also show good consistency with the previous studies. L. J. Kewley & S. L. Ellison (2008) show the MZRs from various metallicity indicators for SDSS galaxies. We find that the MZR from N$_2$O$_2$ is higher than that from N$_2$ by a factor of $\gtrsim 0.2$ dex, which is in good agreement with Fig. 2 in L. J. Kewley & S. L. Ellison (2008) who show that the MZR from N$_2$O$_2$ is higher than that from N$_2$ by a factor of $\sim 0.2$ dex. We attribute the marginal discrepancy to the usage of different versions of N$_2$ indicators.

It is illustrated in Fig. 4 that the MZR from N$_2$O$_2$ has larger scatters compared with that from N$_2$. We attribute this to two potential factors. First, the usage of a constant dust attenuation causes the deviation for the galaxies where [N II]$\lambda$6584 and [O II]$\lambda$3727 are detected while H$\beta$ is not. This is the manifestation of the large dust correction uncertainties considered. Second, the $IZ$-band measurement is subject to shorter exposure time (20 h) compared to that in the $H$ band (100 h).

## 4 SPATIALLY RESOLVED PROPERTIES

In this section, we focus on the galaxy spatially resolved emission line fluxes and metallicities. One of the most interesting chemical properties in IFU era is radial metallicity gradients, a powerful tool to shed light on galaxy baryonic cycles.

We adopt the similar spatial spectrum binning described in Section 3 to enhance the S/N ratios, in each annulus instead of an elliptical aperture. The choice of radial bins is based on the characteristic FWHM of the seeing-dominant PSF ($\sim 0.6$ arcsec) in the KMOS observations. We set three radial bins with widths equal to 0.5 arcsec ($\sim 4.3$ kpc at $z \sim 1.5$), projected on the centres of galaxies (using the position angles and inclination angles listed in Table A1). We integrate the continuum-removed velocity-shift-corrected spectra of the spaxels within each annulus and fit the emission line fluxes (illustrated in Fig. 5). When deriving the metallicities we reject the radial bins where S/Ns of the requested emission lines are lower than 4 (for example, [O II]$\lambda$3727, 9 in annulus B for KURVS_2 in Fig. 5).

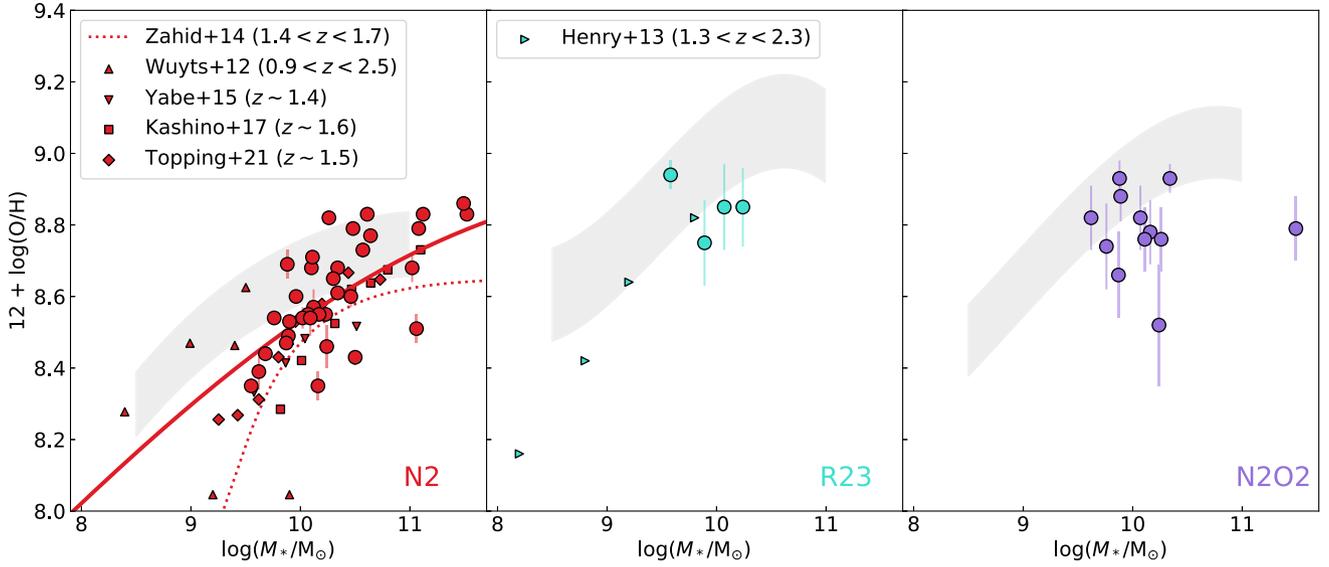

**Figure 4.** Mass–metallicity relations (large circles) for metallicities measured using N₂ (left, F. Bian, L. J. Kewley & M. A. Dopita 2018), R23 (middle, R. Maiolino et al. 2008), and N₂O₂ (right, L. J. Kewley, D. C. Nicholls & R. S. Sutherland 2019) indicators, with the solid curve showing the best fit to the data using the N₂ (left). The typical horizontal error bars are 0.2 dex and considered in the fitting process. The small triangles, squares, and diamonds in the left panel indicate the stellar-mass-binned MZRs from E. Wuyts et al. (2012), K. Yabe et al. (2015), D. Kashino et al. (2017), and M. W. Topping et al. (2021), respectively, while the small triangles in the middle panel show the A. Henry et al. 2013 MZR consisting of 83 emission-line galaxies at $z \in (1.3, 2.3)$ (the error bars are all negligible). The dotted curve in the left panel shows the MZR from H. J. Zahid et al. (2014, at $1.4 < z < 1.7$). The background stripes show the MZRs in the local Universe from the SDSS data reported in L. J. Kewley & S. L. Ellison (2008). All the markers, curves, and stripes in the left panel are recalibrated to F. Bian et al. (2018) N₂ for direct comparison. The best-fitting curves confirm the well-known MZR at $z \sim 1.5$ and are in a good consistency with the previous studies on the N₂ diagnostic at $z \in (0.9, 2.5)$.

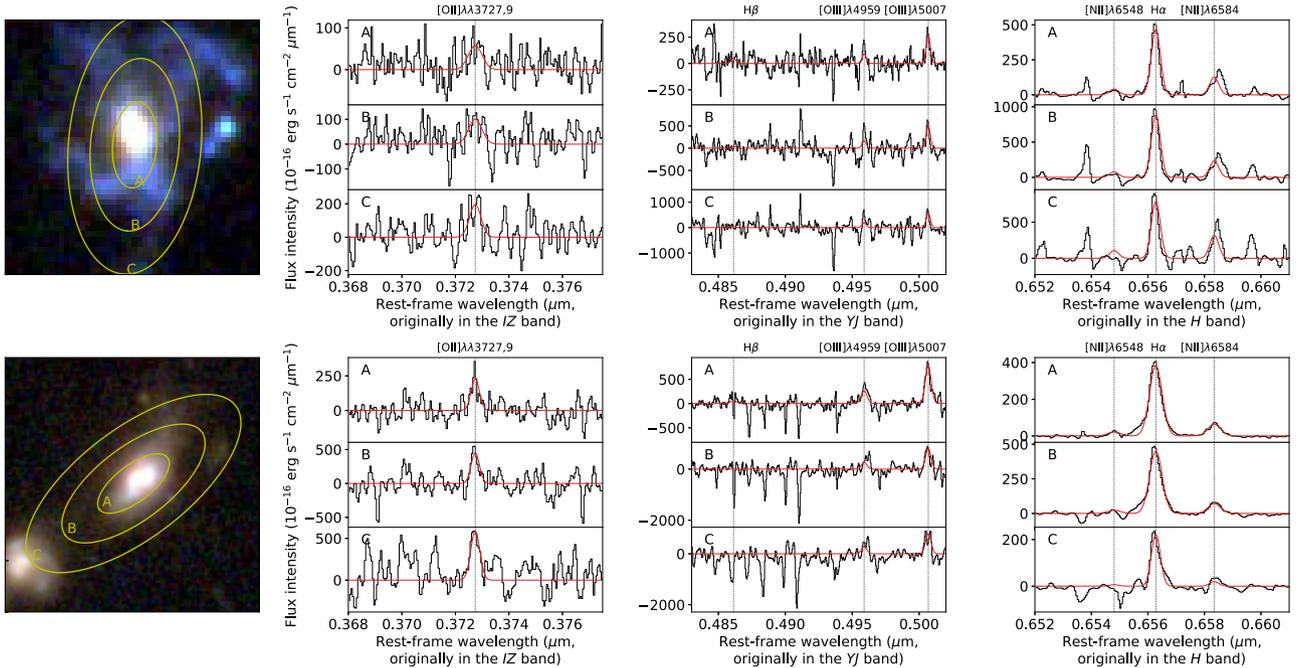

**Figure 5.** Two examples of the imaging and radially binned spectroscopy data of KURVS_2 (upper) and KURVS_34 (lower). The leftmost columns: *HST* (KURVS_2) and *JWST* (KURVS_34) broad-band images. The *HST* image is made from F336W+F435W+F606W as the blue channel, F775W+F814W+F105W as the green channel, and F125W+F160W as the red channel. The *JWST* image is made from F090W+F115W as the blue channel, F150W+F200W+F277W as the green channel, and F356W+F444W as the red channel. The FoV is the same as that of KMOS (2.8 arcsec × 2.8 arcsec). The projected annuli A, B, and C demonstrate our utmost binning edges at 0.5, 1.0, and 1.5 arcsec, respectively. The rest columns: Rest-frame continuum-removed velocity-shift-corrected spectra (steps) and best fits (curves) in each annulus (A, B, and C from top to bottom) and in each band (*IZ*, *YJ*, and *H* from left to right). We do not show the fitting results that have no emission lines with S/N > 4 (the panels without fitted curves).

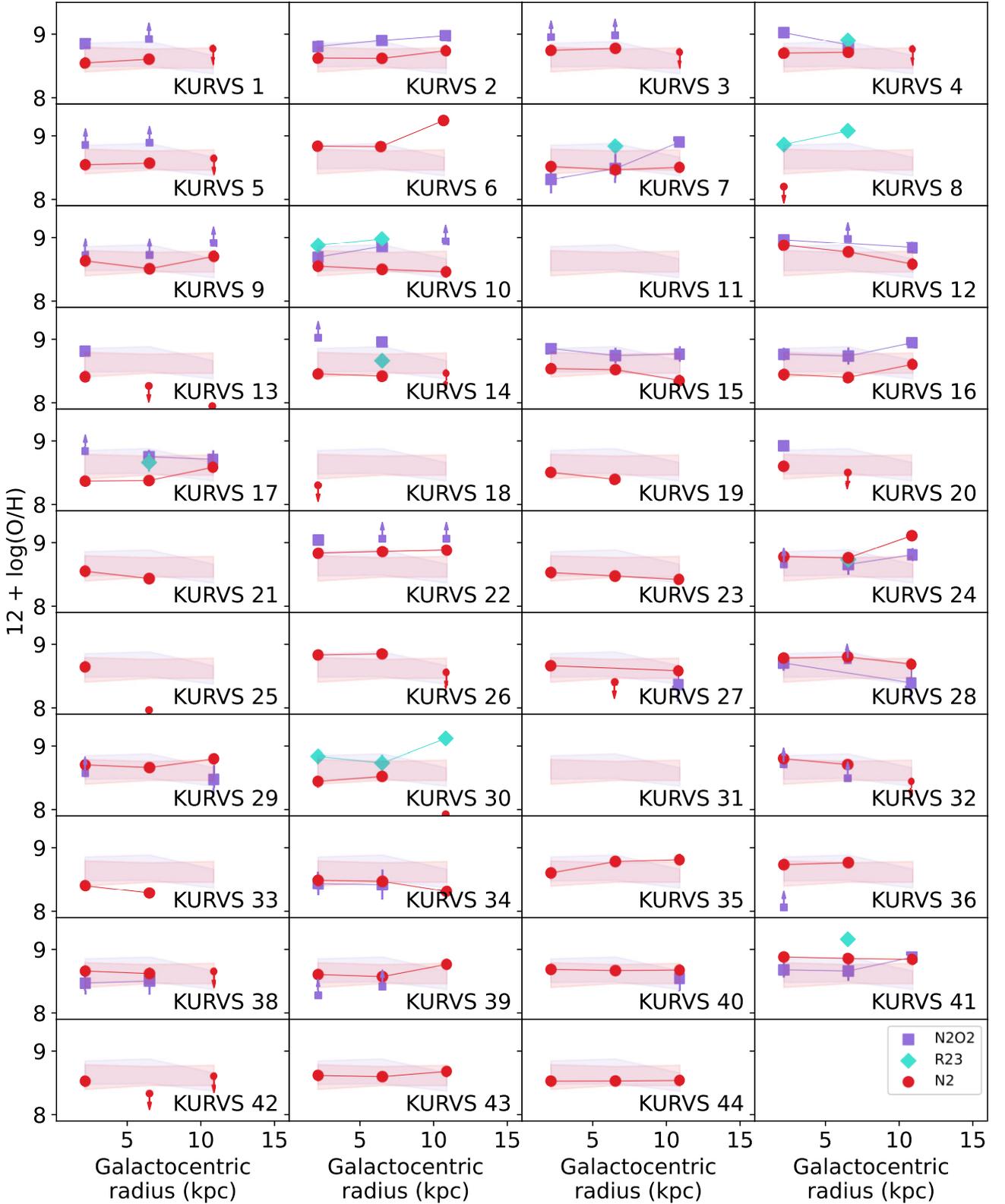

**Figure 6.** Radial metallicity gradients for all the KURVS galaxies in three 0.5 arcsec bins, with circles, diamonds, and squares indicating $N_2$, R23, and $N_2O_2$ metallicity diagnostics (intrinsic, corrected against observing effects), respectively. The large markers show the galaxies with S/Ns of all the emission lines above 4, and the small markers with arrows show the limits of the corresponding line ratios by replacing the line fluxes of which the S/Ns are below 4 by 4× their noises. The background bands represent the 16th and 84th percentiles over the whole sample using $N_2$ (lower) and $N_2O_2$ (higher).

We tried to extend the 0.5 arcsec bin to the further outskirts, and only fewer than ten galaxies showed metallicity gradients from N$_2$ in more than four bins. Thus, we adopt the total bin number of three. To demonstrate that our choice of binning method is robust, we compare changing the bin scheme to two 0.8 arcsec radial bins and to three ∼ 0.6 arcsec radial bins proposed in S. Gillman et al. (2022). The conclusion is that the metallicity gradients are not significantly different using the other two binning methods. The details can be found in Appendix C, and we adopt the choice of three 0.5 arcsec wide radial bins.

As in Section 3.2 we describe integrated dust dereddening, we apply the similar method in the spatially resolved analysis. We stack the spectra of all the galaxies and in each annulus we measure the gas-phase dust attenuation $A_{V, gas}$ from the Balmer decrements of the stacked spectrum of the galaxies. We demonstrate that the values of $A_{V, gas}$ are 0.62 ± 0.16, 0.40 ± 0.20, and 0.00 ± 0.20 for annulus A, B, and C, respectively. This finding agrees well with E. J. Nelson et al. (2016) who report that dust attenuation dominates and has a negative radial gradient in the central regions (< 3 kpc). Thus, in one annulus of a galaxy, if its S/Ns for H$\alpha$ and H$\beta$ are > 4 and its H$\alpha$/H$\beta$ line ratio is > 2.86, we adopt the Balmer-decrement-inferred gas-phase dust attenuation for the annuli; otherwise, we use the dust attenuation from the stacked spectrum in this radial bin.

Beam smearing is one of the most challenging effects that inevitably affects the measured spatially resolved metallicity gradients at high-$z$ (e.g. T. T. Yuan, L. J. Kewley & J. Rich 2013; D. Mast et al. 2014; D. Carton et al. 2017; A. Acharyya et al. 2020). One may expect that a stronger beam smearing effect will erase spatially resolved details thus smoothing measured gradients. Besides, inclination angle also contribute to the effects of beam smearing – the more edge on a galaxy is, the more vulnerable it is against beam smearing, especially along its minor axis. To derive the intrinsic metallicity gradients we follow the correction proposed in S. Gillman et al. (2021) that reconstruct ground truths based on FWHM of seeing and effective radii. The larger inclination angle and the smaller size a galaxy has, the more the measured metallicity deviates from the true value. To summarize, we build a 2D map of the correction coefficient $f_\nabla$(incl., $R_e$) as a function of the inclination angle and effective radius. We estimate the corrected metallicity gradient by multiplying the observed metallicity gradient and $f_\nabla$. For all the galaxies we report a median value of $f_\nabla$ = 1.9 16th and 84th percentiles of $f_\nabla \in$ (1.3, 2.3). We refer readers to S. Gillman et al. (2021) for full details on the scaling relation. From this point on all the radial gradients reported in this work are corrected based on the scaling relation.

We follow the three metallicity indicators (N$_2$, R23, and N$_2$O$_2$) in Section 3.2 and show the individual radial corrected metallicity distributions from all the three diagnostics in Fig. 6. The metallicity gradients are measured as follows. For galaxies that have three available radial bins where the S/Ns of all the required emission line fluxes of the metallicity indicator are above 4, we apply least-$\chi^2$ fitting and obtain the slopes. The uncertainty is estimated using the bootstrapping method mentioned in Section 3.1. For galaxies that have two available radial bins, we adopt straight line slopes between the two metallicity values. For galaxies that have fewer than two available radial bins, we do not report the metallicity gradients. In Table 1, we list the measured metallicity gradients using the pipeline above from the three diagnostics. The majority metallicity gradients are measured from N$_2$O$_2$ and N$_2$, and we focus on these for the rest of this section. In Fig. 7, we show metallicity gradients derived using N$_2$O$_2$ versus those derived using N$_2$. Both illustrate number distributions centred around zero (flat metallicity gradients) with the Pearson correlation between the two metallicity indicators is 0.20 ± 0.44. More than 68 per cent of the data points lie with 0.05 dex kpc$^{-1}$ stripe (light blue background band in Fig. 7), suggesting good agreement.

**Table 1.** Radial metallicity gradients for the KURVS galaxies. (2) Metallicity gradient using N$_2$O$_2$. (3) Metallicity gradient using R23. (4) Metallicity gradient using N$_2$. '∗' marked KURVS_12 is an AGN-dominant galaxy thus the metallicity gradient is subject to diagnostic invalidation.

| Name (1) | $\nabla$(O/H)N$_2$O$_2$ ($10^{-3}$ dex kpc$^{-1}$) (2) | $\nabla$(O/H)R23 ($10^{-3}$ dex kpc$^{-1}$) (3) | $\nabla$(O/H)N$_2$ ($10^{-3}$ dex kpc$^{-1}$) (4) |
|---|---|---|---|
| KURVS_1 | | | 73 ± 46 |
| KURVS_2 | 27 ± 13 | | 19 ± 6 |
| KURVS_3 | | | 11 ± 10 |
| KURVS_4 | | | −2 ± 9 |
| KURVS_5 | | | 13 ± 24 |
| KURVS_6 | | | 81 ± 12 |
| KURVS_7 | 8 ± 99 | | −21 ± 24 |
| KURVS_8 | | 84 ± 33 | |
| KURVS_9 | | | −27 ± 19 |
| KURVS_10 | | | −90 ± 66 |
| KURVS_11 | | | |
| | | | −36 ± 7 |
| KURVS_12∗ | | | |
| KURVS_13 | | | |
| KURVS_14 | | | −14 ± 14 |
| KURVS_15 | −100 ± 129 | | −15 ± 15 |
| KURVS_16 | | | 56 ± 15 |
| KURVS_17 | | | 64 ± 20 |
| KURVS_18 | | | |
| KURVS_19 | | | −74 ± 38 |
| KURVS_20 | | | |
| KURVS_21 | | | −56 ± 18 |
| KURVS_22 | | | 9 ± 5 |
| KURVS_23 | 52 ± 40 | | 36 ± 3 |
| KURVS_24 | 6 ± 52 | | 20 ± 4 |
| KURVS_25 | | | |
| KURVS_26 | | | 25 ± 15 |
| KURVS_27 | | | −219 ± 10 |
| KURVS_28 | | | −58 ± 40 |
| KURVS_29 | | | 39 ± 27 |
| KURVS_30 | | | 124 ± 34 |
| KURVS_31 | | | |
| KURVS_32 | | | −32 ± 9 |
| KURVS_33 | | | |
| KURVS_34 | 34 ± 314 | | −31 ± 14 |
| KURVS_35 | | | 47 ± 10 |
| KURVS_36 | | | −12 ± 8 |
| KURVS_38 | 35 ± 295 | | 23 ± 35 |
| KURVS_39 | | | 34 ± 12 |
| KURVS_40 | | | 1 ± 11 |
| KURVS_41 | −16 ± 42 | | −1 ± 3 |
| KURVS_42 | | | |
| KURVS_43 | | | 17 ± 5 |
| KURVS_44 | | | 10 ± 9 |

In the following discussion, given the consistency of the metallicity gradients using N$_2$O$_2$ and N$_2$, we only focus on the latter for simple statistics. For the 34 galaxies that have a measured metallicity gradient from N$_2$, the typical metallicity gradient for the KURVS sample is 0.01 ± 0.04 dex kpc$^{-1}$ (median values, upper values indicating 84th percentiles and lower values indicating 16th percentiles, respectively).

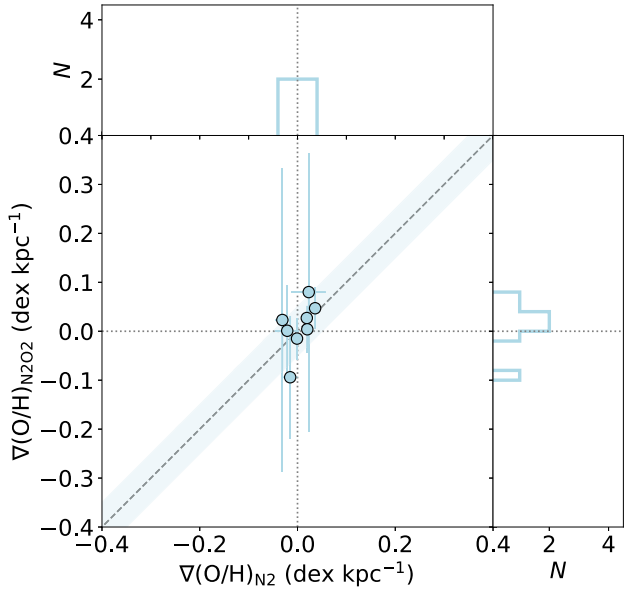

**Figure 7.** Comparison of the measured metallicity gradients using N2 and N2O2 and their number distributions. The dashed line, horizontal dotted line, and vertical dotted line illustrate $y = x$, $y = 0$, and $x = 0$, respectively. The band illustrates the boundaries of $y = x - 0.05$ and $y = x + 0.05$ (in unit of dex kpc$^{-1}$). The Pearson correlation between the two metallicity indicators is $0.20 \pm 0.44$.

### 4.1 How do our results compare with previous studies?

In this section we show how the metallicity gradients obtained above compare with those of other studies on metallicity gradients at similar cosmic epoch of $z \sim 1.5$. The reported metallicity gradients at $z \sim 1.5$ are slightly flatter than those observed in the local Universe ($\sim -0.10$ to $\sim -0.05$ dex kpc$^{-1}$; e.g. H. Poetrodjojo et al. 2018). Our results confirm the findings in the Milky Way and its neighbours that the gas-phase metallicity gradients steepens from $\sim 9$ Gyr ago ($z \sim 1.5$) to present day (L. Stanghellini et al. 2014; L. Magrini et al. 2016). In Fig. 8, we illustrate the metallicity gradient evolution over $z \in (1.1, 1.8)$, including the previous results that use the N2 diagnostic (circles, T. T. Yuan et al. 2011; A. M. Swinbank et al. 2012; N. Leethochawalit et al. 2016; E. Wuyts et al. 2016) and an empirical calibration of [O III]$\lambda$3727, 9/H$\beta$ and [O III]$\lambda$5007/H$\beta$ (diamonds, X. Wang et al. 2017, 2019; M. Curti et al. 2020). We also add the predictions from both analytical chemical evolution models (C. Chiappini et al. 2001; A. Mott et al. 2013; M. Mollá et al. 2019) and numerical simulations (K. Pilkington et al. 2012; B. K. Gibson et al. 2013; X. Ma et al. 2017) as the background lines and stripes.

Our results show good agreement with the existing metallicity gradient measurements using any N2-based indicators (the circles and cyan stars in Fig. 8), which mostly show flat metallicity gradients. Noting that M. Curti et al. (2020) points out that various metallicity indicators will lead to both various absolute metallicity values and various dynamic ranges. In Fig. 8, the results using non-N2 indicators (R23 or Bayesian inference; e.g. X. Wang et al. 2017, 2019; M. Curti et al. 2020, especially non nitrogen emission line included) show statistically wider metallicity gradient ranges but similar median values compared with this work.

The results from numerical simulations and theoretical models can be divided into two groups around $z \sim 1.5$: nearly flat (e.g. C. Chiappini et al. 2001; A. Mott et al. 2013; M. Mollá et al. 2019; M. A. Bellardini et al. 2022; P. B. Tissera et al. 2022; X. Sun et al. 2024) and negative (e.g. G. S. Stinson et al. 2010; A. Acharyya et al. 2025; A. M. Garcia et al. 2025), and in this work the former is favoured. Our results that positive for the half of the sample emphasize that at this cosmic epoch galaxies experience galactic fountains that will be discussed in Section 4.2.

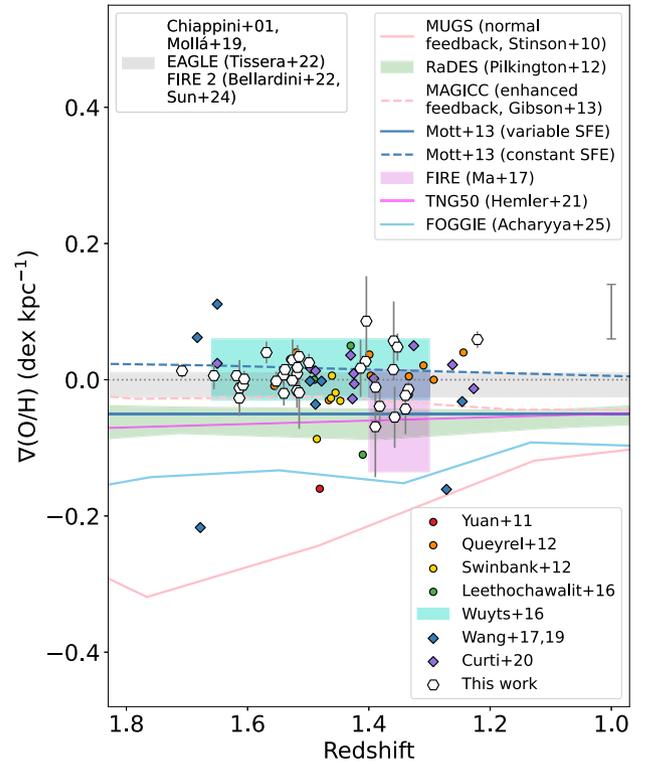

**Figure 8.** Metallicity gradients as a function of redshift in this work, with open hexagons indicating N2. Additional measurements of metallicity gradients at $z \in (1.1, 1.8)$ from previous studies are also included (T. T. Yuan et al. 2011; J. Queyrel et al. 2012; A. M. Swinbank et al. 2012; N. Leethochawalit et al. 2016; E. Wuyts et al. 2016; X. Wang et al. 2017, 2019; M. Curti et al. 2020). The first four studies cited above adopt the N2 metallicity diagnostic proposed by M. Pettini & B. E. J. Pagel (2004). The horizontal dotted line illustrates $y = 0$. The typical error bars of the previous studies are in the lower right corner. We also over-plot results from various analytical models and numerical simulations (C. Chiappini et al. 2001; G. S. Stinson et al. 2010; K. Pilkington et al. 2012; B. K. Gibson et al. 2013; A. Mott et al. 2013; X. Ma et al. 2017; M. Mollá et al. 2019; Z. S. Hemler et al. 2021; M. A. Bellardini et al. 2022; P. B. Tissera et al. 2022; X. Sun et al. 2024; A. Acharyya et al. 2025), out of which we combine some consistent results (C. Chiappini et al. 2001; M. Mollá et al. 2019; M. A. Bellardini et al. 2022; P. B. Tissera et al. 2022; X. Sun et al. 2024) in a grey band to avoid mess. We confirm typically flat metallicity gradients at $z \sim 1.5$ in good consistency with previous studies on observations, theoretical models, and numerical simulations.

### 4.2 What shapes the observed metallicity gradients?

To date a number of analytical and semi-analytical models have been proposed to reconstruct the observed metallicity gradients considering multiple baryonic processes, including inside–out star formation rate gradients, radial gas migration, cosmic gas accretion, and star forming-driven gas outflows (e.g. P. Sharda et al. 2021a). In Fig. 9, we compare the metallicity gradients of our sample with the model proposed by P. Sharda et al. (2021b), as functions of ionized gas rotational velocities ($V_\phi$), velocity dispersions ($\sigma$), and stellar masses.

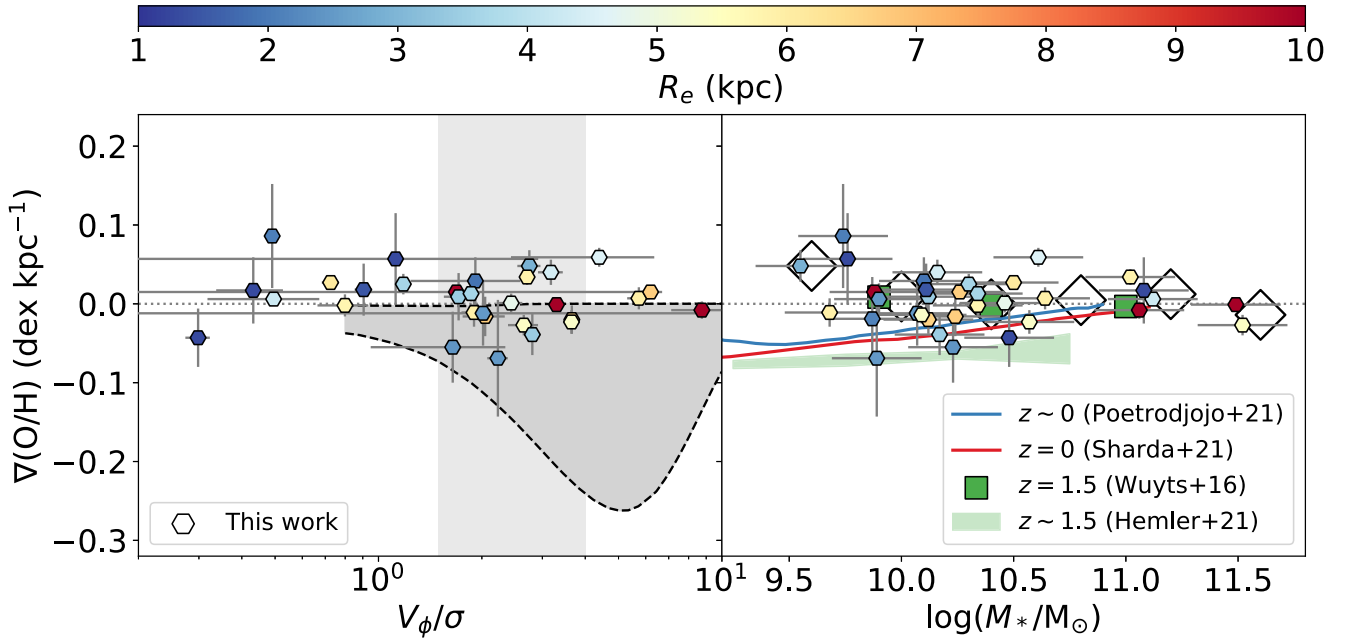

**Figure 9.** Measured metallicity gradients as a function of the ratios of gas rotational velocity and gas velocity dispersion ($V_\phi/\sigma$, left), and stellar masses (right). We colour the data points by effective radii. In the left panel, the dark area indicate the model predicted 'permitted' metallicity gradients proposed in P. Sharda et al. (2021b), between two extremes scenarios of heavy elements being thoroughly mixed and not mixed into the ISM before ejection. The vertical light stripes show the typical $V_\phi/\sigma$ range reported in E. Wisnioski et al. (2015) at $z \sim 1.5$. In the right panel, we plot the binned data in background white diamonds. In both panels, the horizontal dotted lines illustrate $y = 0$. The majority of our sample falls into the 'permitted zone' predicted by the P. Sharda et al. model, and our results well match the observations at the similar redshift range (E. Wuyts et al. 2016).

In the left panel of Fig. 9, the shaded area denotes the 'permitted zone' proposed in P. Sharda et al. (2021b), suggesting that metallicity gradients can only be of negative values when $V_\phi/\sigma$ is intermediate ($\sim 5$), otherwise they would be relatively flat when $V_\phi/\sigma$ is either very large (rotating-dominant) or small (turbulence-dominant). In our sample a majority of the galaxies have $1.5 < V_\phi/\sigma < 4$ (the grey band in the background), on the left side of 'permitted zone'. The 'permitted zone' is enclosed by an upper flat boundary and a lower curvature boundary that represent two extreme cases when the newly produced metals are thoroughly mixed into outflows and the surrounding ISM, respectively. The observed metallicity gradients match well with the typical model predicted metallicity gradients, from $\sim -0.1$ to $0\,\mathrm{dex\,kpc^{-1}}$, suggesting that the self-regulating model can alone explain the observed negative metallicity gradients in this work. However, we see in the left panel of Fig. 9 that inside the 'permitted zone' the massive galaxies locate close to the upper limit of the 'permitted zone'. This might be due to another physical phenomenon that has not been considered in P. Sharda et al. (2021b) but is actually common in massive galaxies, galactic fountains (e.g. F. Fraternali & J. J. Binney 2008; J. K. Werk et al. 2011). This process re-allocates the metals synthesized in the central regions to the outskirts of galaxies, responsible for flattening and even inverting the metallicity gradients. Some observational results also point out that interacting galaxy pairs tend to have flat metallicity gradients (e.g. L. J. Kewley et al. 2010). The higher merger rate that the KURVS galaxies are experiencing at $z \sim 1.5$ than present day can be another possible explanation for the observed flat metallicity gradients.

In the right panel of Fig. 9, we show the metallicity gradients as a function of stellar masses, together with the reference observational results reported in H. Poetrodjojo et al. (2021) and E. Wuyts et al. (2016), simulated results shown in Z. S. Hemler et al. (2021) and the theoretical model proposed in P. Sharda et al. (2021a). Our results agree well with those in E. Wuyts et al. (2016) at all stellar masses, and with those in H. Poetrodjojo et al. (2021) at $\log(M_*/M_\odot) > 10$. The stellar mass dependence of gas-phase metallicity gradient at $z \sim 1.5$ does not exhibit significant deviation from that observed in the local Universe within the range of $10 < \log(M_*/M_\odot) < 11$, suggesting that the processes governing baryonic regulation in the local Universe remain applicable at high-$z$ (F. Belfiore et al. 2017). Observations of nearby galaxies indicate that massive galaxies typically have flat metallicity gradients that can be attributed to metal yield saturation in their central regions. An alternative explanation involves the direct accretion of metal-poor gas from the cosmic web preferentially affecting the outer regions of galaxies. The radial metallicity profiles of KURVS_12 and KURVS_26 in Fig. 6 are consistent with this scenario. At the low-mass end, however, our results reveal that a majority of the sample exhibits positive metallicity gradients, a trend reported by E. Wuyts et al. (2016). In the local Universe star-forming dwarf galaxies are found to have flat or positive metallicity gradients (e.g. M. Grossi et al. 2020). The critical stellar mass below which positive metallicity gradients become prevalent is $\sim 1.5 \times 10^9\,M_\odot$ for local galaxies (F. Belfiore et al. 2017). In contrast, our findings indicate that this stellar-mass-dependent transition occurs at a higher stellar mass of $\sim 10^{10}\,M_\odot$. This shift may reflect the fact that galaxy discs at this epoch are not yet fully settled and are undergoing clumpy star formation, where the competition of gas inflows and outflows shapes the metallicity distribution (M. Grossi et al. 2020).

In summary, we ascribe our observed negative metallicity gradients to more turbulent discs ($V_\phi/\sigma < 5$) that thoroughly mix the metal-rich gas (P. Sharda et al. 2021b). The observed positive metallicity gradients may be due to the factors that are not included in theoretical models: galactic fountains mostly found in massive galaxies that flatten the metallicity gradients (F. Fraternali & J. J.

Binney [2008](#); J. K. Werk et al. [2011](#)) and higher merger rates at $z \sim 1.5$ that flatten the metallicity gradients (L. J. Kewley et al. [2010](#)).

## 5 CONCLUSIONS

We have analysed a sample of 43 star-forming galaxies at $z \sim 1.5$ observed with KMOS on the VLT. The observations conducted in the *IZ*, *YJ*, and *H* bands provide both integrated and spatially resolved rest-frame optical nebular emission line properties, based on which we derive the corresponding ISM oxygen abundances. In this paper, we report the gas-phase metallicities from the [N II]/[O II] strong-line calibrations that traces primarily N/O ratios and are relatively insensitive against ionization parameter and ISM pressure. We summarize the main conclusions drawn in this paper as follows:

(i) We confirm that the integrated properties of the KURVS sample in the BPT diagram agree well with the previous studies at the similar cosmic epoch. We report the well-known mass-metallicity relations at $z \sim 1.5$ traditionally from the [N II]$\lambda6584$/H$\alpha$ (N$_2$) and ([O II]$\lambda\lambda3727$, 9+[O III]$\lambda\lambda4959$, 5007)/H$\beta$ (R23) metallicity indicators, and for the first time from the [N II]$\lambda6584$/[O II]$\lambda\lambda3727$, 9 (N$_2$O$_2$) metallicity indicator, one of the most reliable metallicity diagnostics.

(ii) We report the spatially resolved metallicity gradients from the metallicity indicators above, considering dust effects. The metallicity gradients measured from N$_2$O$_2$ agree well with those measured from N$_2$ for a subsample of eight galaxies. We take the metallicity gradients from N$_2$ as our default choice. We report the metallicity gradients for a subset of 34 galaxies. Our sample typically shows nearly flat metallicity gradients ($0.01 \pm 0.04$ dex kpc$^{-1}$). The metallicity gradient measurements are corrected for the flattening effect from PSF, and robust against the choice of binning scheme.

(iii) The observed metallicity gradients agree well with the previous studies on both observations and simulations at this epoch. We ascribe the observed flat metallicity gradients to turbulent discs, higher merger rates, and galactic fountains.

## ACKNOWLEDGEMENTS

ZL, AMS, and IS acknowledge the Science and Technology Facilities Council (STFC) consolidated grant ST/X001075/1. AP acknowledges the STFC consolidated grants ST/T000244/1 and ST/P000541/1. SG acknowledges the support of the Cosmic Dawn Center of Excellence funded by the Danish National Research Foundation under the grant 140. DJT acknowledges the STFC studentship (ST/X508354/1). LC acknowledges support from the Australian Research Council Discovery Project funding scheme (DP210100337). CH acknowledges funding from a United Kingdom Research and Innovation grant (MR/V022830/1). EI and JM gratefully acknowledge financial support from ANID – MILENIO – NCN2024_112 and ANID FONDECYT Regular 1221846.

## DATA AVAILABILITY

The data and the source code underlying this article are available at https://github.com/zidianjun/kurvs.

## REFERENCES

Acharyya A., Krumholz M. R., Federrath C., Kewley L. J., Goldbaum N. J., Sharp R., 2020, MNRAS, 495, 3819
Acharyya A. et al., 2025, ApJ, 979, 129
Andrews B. H., Martini P., 2013, ApJ, 765, 140
Baldwin J. A., Phillips M. M., Terlevich R., 1981, PASP, 93, 5
Battisti A. J. et al., 2022, MNRAS, 513, 4431
Belfiore F. et al., 2017, MNRAS, 469, 151
Bellardini M. A., Wetzel A., Loebman S. R., Bailin J., 2022, MNRAS, 514, 4270
Bian F., Kewley L. J., Dopita M. A., 2018, ApJ, 859, 175
Brinchmann J., Charlot S., White S. D. M., Tremonti C., Kauffmann G., Heckman T., Brinkmann J., 2004, MNRAS, 351, 1151
Bundy K. et al., 2015, ApJ, 798, 7
Carton D. et al., 2017, MNRAS, 468, 2140
Cataldi E. et al., 2025, preprint (arXiv:2504.03839)
Chemerynska I. et al., 2024, ApJ, 976, L15
Chiappini C., Matteucci F., Gratton R., 1997, ApJ, 477, 765
Chiappini C., Matteucci F., Romano D., 2001, ApJ, 554, 1044
Croom S. M. et al., 2012, MNRAS, 421, 872
da Cunha E., Charlot S., Elbaz D., 2008, MNRAS, 388, 1595
da Cunha E. et al., 2015, ApJ, 806, 110
Curti M. et al., 2020, MNRAS, 492, 821
Curti M. et al., 2024, A&A, 684, A75
Domínguez A. et al., 2013, ApJ, 763, 145
Dopita M. A., Kewley L. J., Sutherland R. S., Nicholls D. C., 2016, Ap&SS, 361, 61 (D16)
Emsellem E. et al., 2022, A&A, 659, A191
Erb D. K., Steidel C. C., Shapley A. E., Pettini M., Reddy N. A., Adelberger K. L., 2006, ApJ, 646, 107
Erroz-Ferrer S. et al., 2019, MNRAS, 484, 5009
Fraternali F., Binney J. J., 2008, MNRAS, 386, 935
Garcia A. M. et al., 2025, ApJ, 989, 147
Gibson B. K., Pilkington K., Brook C. B., Stinson G. S., Bailin J., 2013, A&A, 554, A47
Gillman S. et al., 2020, MNRAS, 492, 1492
Gillman S. et al., 2021, MNRAS, 500, 4229
Gillman S. et al., 2022, MNRAS, 512, 3480
Grogin N. A. et al., 2011, ApJS, 197, 35
Grossi M., García-Benito R., Cortesi A., Gonçalves D. R., Gonçalves T. S., Lopes P. A. A., Menéndez-Delmestre K., Telles E., 2020, MNRAS, 498, 1939
Guo Y. et al., 2013, ApJS, 207, 24
Hayden-Pawson C. et al., 2022, MNRAS, 512, 2867
Hemler Z. S. et al., 2021, MNRAS, 506, 3024
Henry A. et al., 2013, ApJ, 776, L27
Henry R. B. C., Worthey G., 1999, PASP, 111, 919
Ho I. T., Kudritzki R.-P., Kewley L. J., Zahid H. J., Dopita M. A., Bresolin F., Rupke D. S. N., 2015, MNRAS, 448, 2030
Jones T., Sanders R., Roberts-Borsani G., Ellis R. S., Laporte N., Treu T., Harikane Y., 2020, ApJ, 903, 150
Juneau S. et al., 2014, ApJ, 788, 88
Kashino D. et al., 2017, ApJ, 835, 88
Kashino D. et al., 2019, ApJS, 241, 10
Kauffmann G. et al., 2003a, MNRAS, 341, 33
Kauffmann G. et al., 2003b, MNRAS, 346, 1055
Kewley L. J., Dopita M. A., 2002, ApJS, 142, 35
Kewley L. J., Ellison S. L., 2008, ApJ, 681, 1183
Kewley L. J., Dopita M. A., Sutherland R. S., Heisler C. A., Trevena J., 2001, ApJ, 556, 121
Kewley L. J., Rupke D., Zahid H. J., Geller M. J., Barton E. J., 2010, ApJ, 721, L48
Kewley L. J., Nicholls D. C., Sutherland R. S., 2019, ARA&A, 57, 511
Kobulnicky H. A., Kewley L. J., 2004, ApJ, 617, 240
Kreckel K. et al., 2019, ApJ, 887, 80
Laseter I. H. et al., 2024, A&A, 681, A70
Lee H., Skillman E. D., Cannon J. M., Jackson D. C., Gehrz R. D., Polomski E. F., Woodward C. E., 2006, ApJ, 647, 970
Leethochawalit N., Jones T. A., Ellis R. S., Stark D. P., Richard J., Zitrin A., Auger M., 2016, ApJ, 820, 84
Lian J. H., Li J. R., Yan W., Kong X., 2015, MNRAS, 446, 1449

López-Cobá C. et al., 2020, AJ, 159, 167

Ma X., Hopkins P. F., Feldmann R., Torrey P., Faucher-Giguère C.-A., Kereš D., 2017, MNRAS, 466, 4780

Magrini L., Coccato L., Stanghellini L., Casasola V., Galli D., 2016, A&A, 588, A91

Maiolino R. et al., 2008, A&A, 488, 463

Mannucci F., Cresci G., Maiolino R., Marconi A., Gnerucci A., 2010, MNRAS, 408, 2115

Marino R. A. et al., 2013, A&A, 559, A114

Mast D. et al., 2014, A&A, 561, A129

Mathavu J. et al., 2023, ApJ, 949, L11

Molina J., Ibar E., Swinbank A. M., Sobral D., Best P. N., Smail I., Escala A., Cirasuolo M., 2017, MNRAS, 466, 892

Mollá M., Ferrini F., Díaz A. I., 1997, ApJ, 475, 519

Mollá M., Díaz Á. I., Cavichia O., Gibson B. K., Maciel W. J., Costa R. D. D., Ascasibar Y., Few C. G., 2019, MNRAS, 482, 3071

Mott A., Spitoni E., Matteucci F., 2013, MNRAS, 435, 2918

Muzzin A. et al., 2013, ApJS, 206, 8

Naab T., Ostriker J. P., 2006, MNRAS, 366, 899

Nakajima K., Ouchi M., Isobe Y., Harikane Y., Zhang Y., Ono Y., Umeda H., Oguri M., 2023, ApJS, 269, 33

Nelson E. J. et al., 2016, ApJ, 817, L9

Osterbrock D. E., 1989, Astrophysics of Gaseous Nebulae and Active Galactic Nuclei. University Science Books

Pagel B. E. J., Edmunds M. G., Smith G., 1980, MNRAS, 193, 219

Pérez-Montero E., 2014, MNRAS, 441, 2663

Pettini M., Pagel B. E. J., 2004, MNRAS, 348, L59

Pilkington K. et al., 2012, A&A, 540, A56

Planck Collaboration XVI, 2014, A&A, 571, A16

Poetrodjojo H. et al., 2018, MNRAS, 479, 5235

Poetrodjojo H. et al., 2021, MNRAS, 502, 3357

Puglisi A. et al., 2023, MNRAS, 524, 2814

Queyrel J. et al., 2012, A&A, 539, A93

Sánchez-Menguiano L. et al., 2017, A&A, 603, A113

Sánchez S. F. et al., 2012, A&A, 538, A8

Sánchez S. F. et al., 2014, A&A, 563, A49

Sanders R. L., Shapley A. E., Topping M. W., Reddy N. A., Brammer G. B., 2024, ApJ, 962, 24

Scholte D. et al., 2025, MNRAS, 540, 1800

Schreiber C. et al., 2015, A&A, 575, A74

Searle L., 1971, ApJ, 168, 327

Shapley A. E. et al., 2015, ApJ, 801, 88

Shapley A. E. et al., 2025, ApJ, 980, 242

Sharda P., Krumholz M. R., Wisnioski E., Forbes J. C., Federrath C., Acharyya A., 2021a, MNRAS, 502, 5935

Sharda P., Wisnioski E., Krumholz M. R., Federrath C., 2021b, MNRAS, 506, 1295

Shields G. A., 1974, ApJ, 193, 335

Spitoni E., Matteucci F., Marcon-Uchida M. M., 2013, A&A, 551, A123

Spitoni E., Romano D., Matteucci F., Ciotti L., 2015, ApJ, 802, 129

Stanghellini L., Magrini L., Casasola V., Villaver E., 2014, A&A, 567, A88

Steidel C. C. et al., 2014, ApJ, 795, 165

Stinson G. S., Bailin J., Couchman H., Wadsley J., Shen S., Nickerson S., Brook C., Quinn T., 2010, MNRAS, 408, 812

Strom A. L., Steidel C. C., Rudie G. C., Trainor R. F., Pettini M., Reddy N. A., 2017, ApJ, 836, 164

Sun X. et al., 2024, ApJ, 986, 179

Swinbank A. M., Sobral D., Smail I., Geach J. E., Best P. N., McCarthy I. G., Crain R. A., Theuns T., 2012, MNRAS, 426, 935

Tiley A. L. et al., 2021, MNRAS, 506, 323

Tinsley B. M., 1980, Fund. Cosmic Phys., 5, 287

Tissera P. B., Rosas-Guevara Y., Sillero E., Pedrosa S. E., Theuns T., Bignone L., 2022, MNRAS, 511, 1667

Topping M. W. et al., 2021, MNRAS, 506, 1237

Tremonti C. A. et al., 2004, ApJ, 613, 898

Troncoso P. et al., 2014, A&A, 563, A58

Vila-Costas M. B., Edmunds M. G., 1992, MNRAS, 259, 121

Vilchez J. M., Esteban C., 1996, MNRAS, 280, 720

Wang X. et al., 2017, ApJ, 837, 89

Wang X. et al., 2019, ApJ, 882, 94

Werk J. K., Putman M. E., Meurer G. R., Santiago-Figueroa N., 2011, ApJ, 735, 71

Wisnioski E. et al., 2015, ApJ, 799, 209

Wisnioski E. et al., 2019, ApJ, 886, 124

Wuyts E., Rigby J. R., Sharon K., Gladders M. D., 2012, ApJ, 755, 73

Wuyts E. et al., 2016, ApJ, 827, 74

Wuyts S. et al., 2013, ApJ, 779, 135

Yabe K. et al., 2015, PASJ, 67, 102

Yuan T. T., Kewley L. J., Swinbank A. M., Richard J., Livermore R. C., 2011, ApJ, 732, L14

Yuan T. T., Kewley L. J., Rich J., 2013, ApJ, 767, 106

Zahid H. J., Kewley L. J., Bresolin F., 2011, ApJ, 730, 137

Zahid H. J. et al., 2014, ApJ, 792, 75

Zaritsky D., Kennicutt Robert C. J., Huchra J. P., 1994, ApJ, 420, 87

## APPENDIX A: BASIC INTEGRATED PROPERTIES OF THE KURVS SAMPLE

In this appendix we show the integrated properties (coordinates, stellar masses, star formation rates, and deprojection information) of the KURVS sample in Table A1. We also show the photometric data and integrated spectra of the galaxies within a circular aperture with a deprojected radius of 1.2 arcsec ($\sim 10$ kpc at $z \sim 1.5$) and corrected for galaxy rotational velocities for all the KURVS galaxies in four figures from Fig. A1 to Fig. A4. The photometric images are from *JWST* where available, otherwise from *HST*. The spectra are truncated to illustrate only the strong emission lines in three independent KMOS *NIR* bands, from short wavelengths to long wavelengths, $IZ$, $YJ$, and $H$. The emission lines are [O II]$\lambda\lambda3727$, 9 in the $IZ$ band, H$\beta$, [O III]$\lambda4959$, and [O III]$\lambda5007$ in the $YJ$ band, [N II]$\lambda6548$, H$\alpha$, and [N II]$\lambda6584$ in the $H$ band. The wavelengths are shifted to rest frame so that they share the same $x$-axis. To shorten the names we only show the ID numbers for the target galaxies.

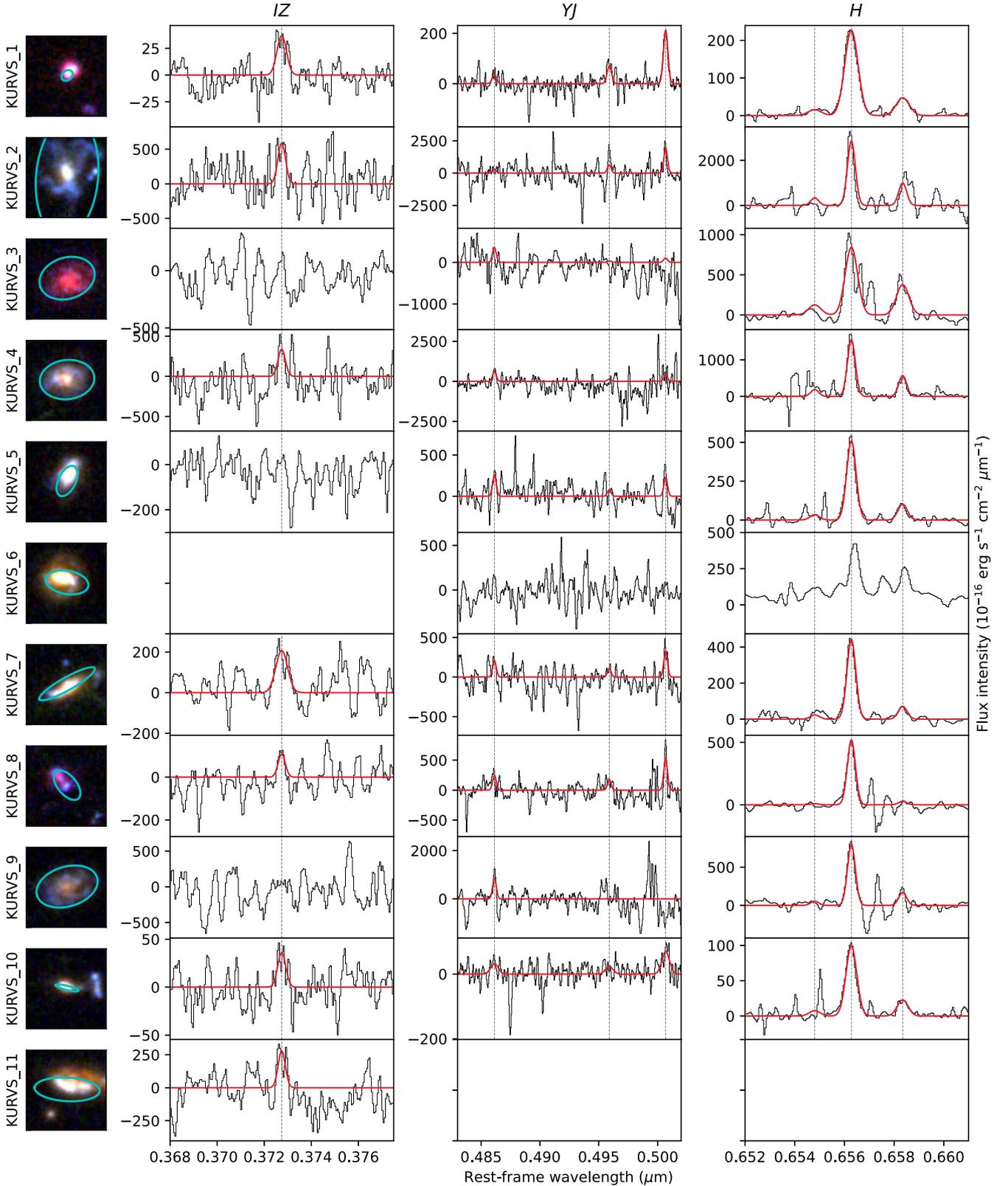

**Figure A1.** The photometric and spectroscopic data for the galaxies in the KURVS sample. The leftmost column: *JWST* or *HST* broad-band images with their KURVS ID numbers labelled on the left, similar to the left column of Fig. 5. The ellipses show the apertures with radii of $1.5R_e$. The rest columns: From left to right, the steps show the rest-frame integrated emission line spectra (continua removed and corrected for galaxy rotational velocity shifts) for the KURVS galaxies in the $IZ$, $YJ$, and $H$ bands, respectively. The blank panels demonstrate the absence of observations. The curves are the best least-$\chi^2$ Gaussian fits and we do not show the fitting results that have no emission lines with S/N > 4. The vertical dashed lines indicate the emission lines of [O II]$\lambda\lambda3727, 9$ (doublet spectroscopically unresolved, originally in the $IZ$ band), H$\beta$+[O III]$\lambda4959$+[O III]$\lambda5007$ (originally in the $YJ$ band), and [N II]$\lambda6548$+H$\alpha$+[N II]$\lambda6584$ (originally in the $H$ band).

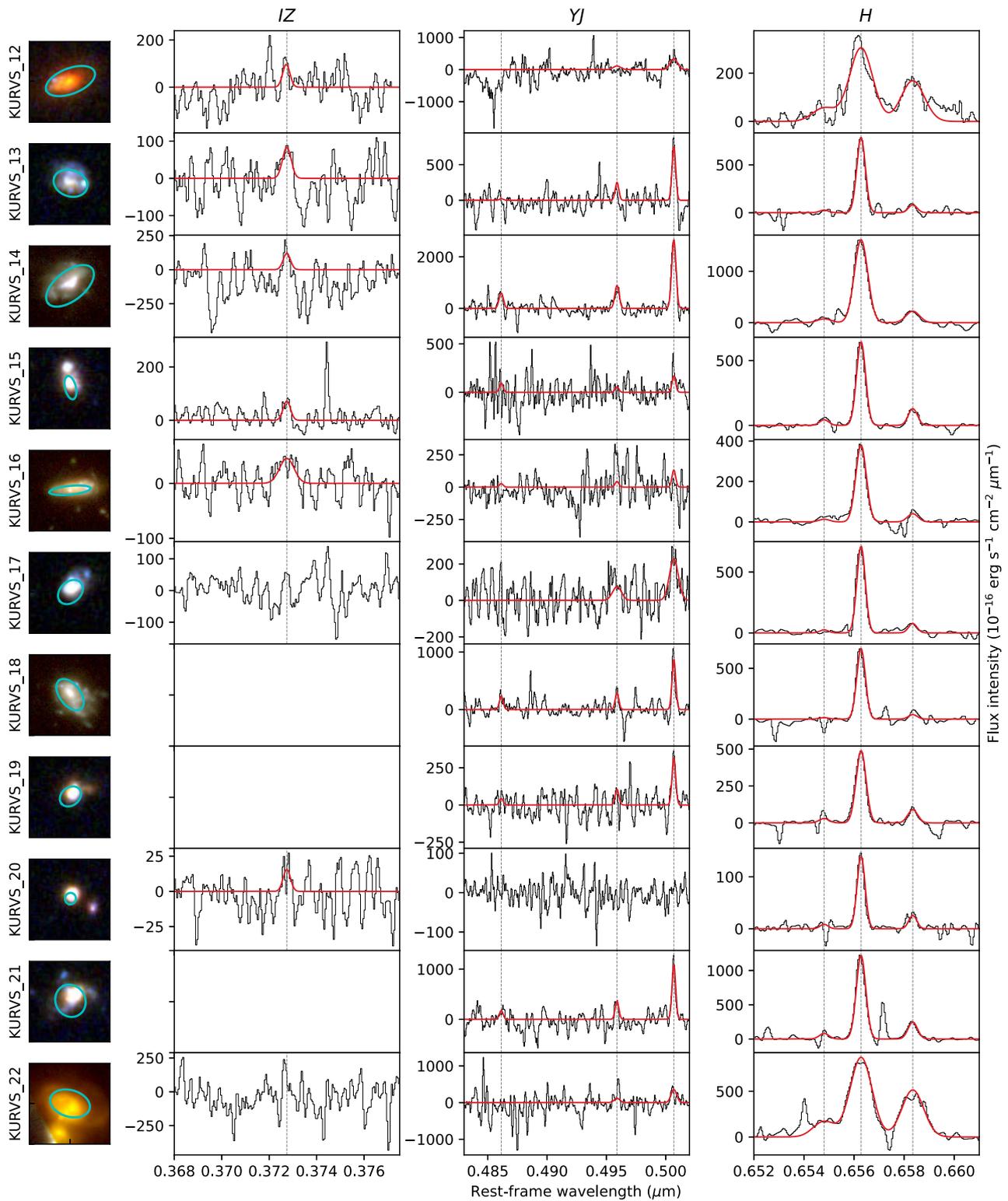

**Figure A2.** Continued, as in Fig. A1 for the KURVS galaxies from KURVS_12 to KURVS_22.

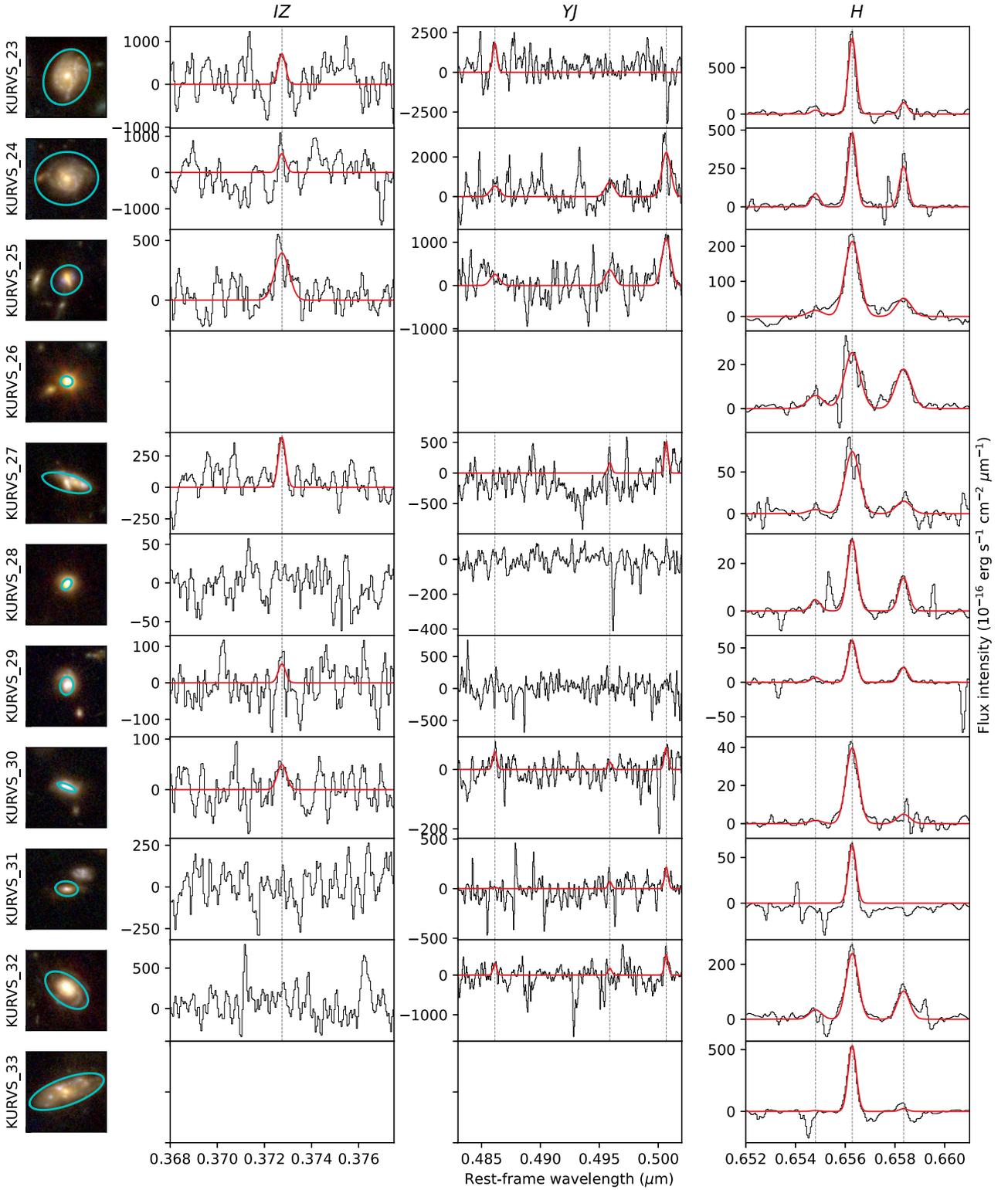

**Figure A3.** Continued, as in Fig. A1 for the KURVS galaxies from KURVS_23 to KURVS_33.

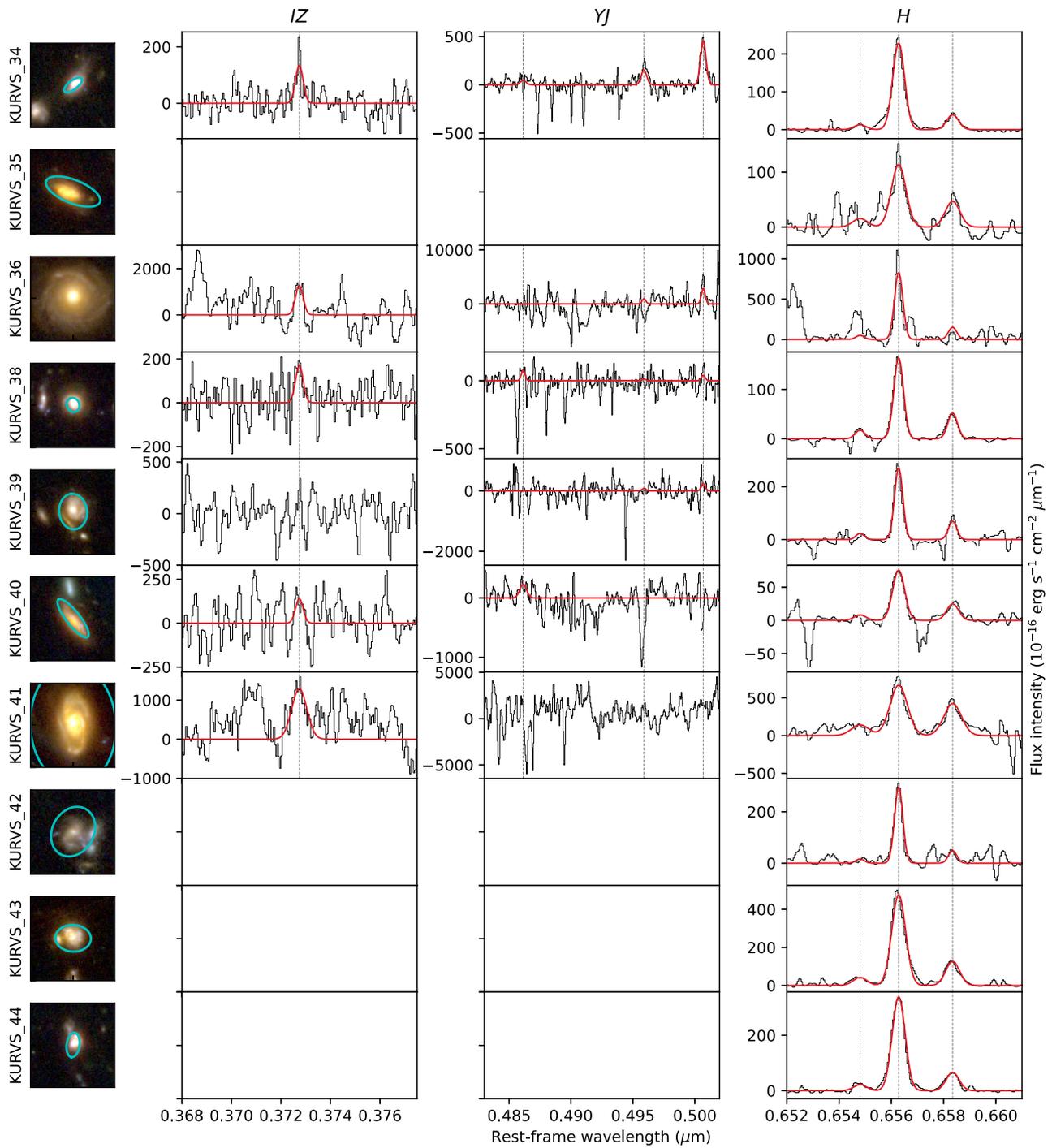

**Figure A4.** Continued, as in Fig. A1 for the KURVS galaxies from KURVS_34 to KURVS_44.

**Table A1.** (1) Galaxy ID from KURVS (#37 is not detected); (2) name of the target; (3) and (4) sky coordinates; (5) spectroscopic redshift from H$\alpha$ emission lines; (6) stellar mass estimated using *HST* broad images (with error ∼ 0.2 dex); (7) star formation rate estimated using *HST* broad images (with error ∼ 0.05 dex); (8) and (9) position angle and inclination angle (0° = face-on and 90° = edge-on) estimated using *HST* images; and (10) effective radius.

| KURVS ID | Name | RA (hh:mm:ss) | Dec. (dd:mm:ss) | $z_{spec,H\alpha}$ | $\log(M_*/M_\odot)$ | SFR ($M_\odot$ yr$^{-1}$) | PA (°) | incl. (°) | $R_e$ (") |
|---|---|---|---|---|---|---|---|---|---|
| (1) | (2) | (3) | (4) | (5) | (6) | (7) | (8) | (9) | (10) |
| 1 | J03321735−27535283 | 03:32:17.35 | −27:53:52.83 | 1.3590 | 9.76 | 36 | −49 | 38 | 0.15 |
| 2 | J03321500−27530237 | 03:32:15.00 | −27:53:02.37 | 1.3597 | 9.88 | 12 | −5 | 59 | 1.49 |
| 3 | J03322210−27524495 | 03:32:22.10 | −27:52:44.95 | 1.5409 | 10.64 | 34 | −73 | 43 | 0.69 |
| 4 | J03322018−27523834 | 03:32:20.18 | −27:52:38.34 | 1.5527 | 10.34 | 62 | −84 | 44 | 0.67 |
| 5 | J03321378−27520273 | 03:32:13.78 | −27:52:02.73 | 1.5182 | 10.12 | 33 | −26 | 58 | 0.40 |
| 6 | J03321405−27512440 | 03:32:14.05 | −27:51:24.40 | 1.2209 | 10.61 | 35 | 80 | 57 | 0.53 |
| 7 | J03321123−27510710 | 03:32:11.23 | −27:51:07.10 | 1.5185 | 10.24 | 14 | −61 | 78 | 0.78 |
| 8 | J03320683−27505537 | 03:32:06.83 | −27:50:55.37 | 1.5393 | 9.58 | 18 | 40 | 57 | 0.43 |
| 9 | J03322152−27504053 | 03:32:21.52 | −27:50:40.53 | 1.5402 | 10.12 | 10 | −73 | 51 | 0.77 |
| 10 | J03322992−27503191 | 03:32:29.92 | −27:50:31.91 | 1.3893 | 9.89 | 4 | 76 | 80 | 0.29 |
| 11 | J03323403−27502882 | 03:32:34.03 | −27:50:28.82 | 1.3827 | 10.68 | 142 | 87 | 69 | 0.80 |
| 12 | J03323154−27502868 | 03:32:31.54 | −27:50:28.68 | 1.6136 | 11.52 | 38 | −69 | 61 | 0.63 |
| 13 | J03321250−27502059 | 03:32:12.50 | −27:50:20.59 | 1.3337 | 9.62 | 39 | 71 | 41 | 0.42 |
| 14 | J03323737−27501362 | 03:32:37.37 | −27:50:13.62 | 1.3890 | 9.68 | 30 | −52 | 59 | 0.69 |
| 15 | J03321694−27500406 | 03:32:16.94 | −27:50:04.06 | 1.6133 | 10.07 | 43 | 13 | 63 | 0.29 |
| 16 | J03323710−27494094 | 03:32:37.10 | −27:49:40.94 | 1.5689 | 10.16 | 25 | −84 | 78 | 0.51 |
| 17 | J03321112−27493841 | 03:32:11.12 | −27:49:38.41 | 1.3528 | 9.55 | 17 | −48 | 42 | 0.34 |
| 18 | J03323264−27484872 | 03:32:32.64 | −27:48:48.72 | 1.3416 | 9.82 | 17 | 37 | 54 | 0.44 |
| 19 | J03321979−27483912 | 03:32:19.79 | −27:48:39.12 | 1.3574 | 10.23 | 18 | −51 | 39 | 0.28 |
| 20 | J03322503−27471818 | 03:32:25.03 | −27:47:18.18 | 1.3564 | 10.02 | 63 | −10 | 18 | 0.14 |
| 21 | J03323143−27513748 | 03:32:31.43 | −27:51:37.48 | 1.3821 | 10.17 | 21 | 27 | 25 | 0.40 |
| 22 | J03323774−27500039 | 03:32:37.74 | −27:50:00.39 | 1.6182 | 11.12 | 101 | 74 | 49 | 0.50 |
| 23 | J10003429+02222370 | 10:00:34.29 | +02:22:23.70 | 1.4054 | 10.50 | 35 | −24 | 40 | 0.71 |
| 24 | J10003544+02205961 | 10:00:35.44 | +02:20:59.61 | 1.5384 | 10.26 | 31 | −87 | 33 | 0.78 |
| 25 | J10003333+02222415 | 10:00:33.33 | +02:22:24.15 | 1.3686 | 9.96 | 4 | −53 | 32 | 0.40 |
| 26 | J10003437+02205566 | 10:00:34.37 | +02:20:55.66 | 1.4136 | 11.08 | 129 | 84 | 32 | 0.15 |
| 27 | J10003271+02211934 | 10:00:32.71 | +02:21:19.34 | 1.3348 | 10.09 | 12 | 75 | 70 | 0.62 |
| 28 | J10002643+02192844 | 10:00:26.43 | +02:19:28.44 | 1.3395 | 10.48 | 18 | −38 | 46 | 0.15 |
| 29 | J10002034+02211926 | 10:00:20.34 | +02:21:19.26 | 1.5258 | 10.10 | 29 | −6 | 41 | 0.22 |
| 30 | J10002221+02192810 | 10:00:22.21 | +02:19:28.10 | 1.4041 | 9.74 | 19 | 68 | 72 | 0.24 |
| 31 | J10002413+02190927 | 10:00:24.13 | +02:19:09.27 | 1.5185 | 9.49 | 8 | 84 | 50 | 0.28 |
| 32 | J10002066+02181580 | 10:00:20.66 | +02:18:15.80 | 1.3396 | 10.57 | 25 | 52 | 57 | 0.62 |
| 33 | J10001898+02180628 | 10:00:18.98 | +02:18:06.28 | 1.4636 | 10.24 | 30 | −70 | 70 | 0.97 |
| 34 | J10002764+02182477 | 10:00:27.64 | +02:18:24.77 | 1.5149 | 9.87 | 25 | −53 | 65 | 0.25 |
| 35 | J10002630+02162743 | 10:00:26.30 | +02:16:27.43 | 1.5150 | 11.02 | 34 | 68 | 66 | 0.68 |
| 36 | J10003355+02164671 | 10:00:33.55 | +02:16:46.71 | 1.6076 | 11.06 | 10 | −57 | 17 | 1.49 |
| 38 | J10003255+02175601 | 10:00:32.55 | +02:17:56.01 | 1.5177 | 10.11 | 53 | 23 | 29 | 0.17 |
| 39 | J10004254+02184696 | 10:00:42.54 | +02:18:46.96 | 1.4987 | 10.30 | 27 | 7 | 39 | 0.42 |
| 40 | J10003941+02180390 | 10:00:39.41 | +02:18:03.90 | 1.6056 | 10.46 | 17 | 38 | 73 | 0.56 |
| 41 | J10004503+02192099 | 10:00:45.03 | +02:19:20.99 | 1.5262 | 11.49 | 41 | 3 | 45 | 1.43 |
| 42 | J10004251+02204899 | 10:00:42.51 | +02:20:48.99 | 1.5002 | 9.72 | 15 | −29 | 37 | 0.61 |
| 43 | J10003687+02213020 | 10:00:36.87 | +02:21:30.20 | 1.7082 | 10.34 | 49 | 87 | 42 | 0.42 |
| 44 | J10003631+02211750 | 10:00:36.31 | +02:21:17.50 | 1.6556 | 9.90 | 42 | −8 | 58 | 0.29 |

## APPENDIX B: COMPARISON OF METALLICITY INDICATORS

Our data cover a wide wavelength range thus making it possible to investigate multiple metallicity indicators. In addition to the three metallicity diagnostics mentioned in Section 3.2 (N2, R23, and N2O2) we consider [O III]λ5007×H α/[N II]λ6584/H β (O3N2; F. Bian et al. 2018) and the diagnostic proposed by M. A. Dopita et al. (2016, hereafter D16). We show the corner plot Fig. B1 to compare the metallicities from the five indicators. We confirm the steeper slope of D16 metallicities compared with N2 metallicities, reported by S. Gillman et al. (2022). The crossing point is at $12 + \log(\text{O/H}) \sim 8.4$, above which D16 returns a larger value than N2.

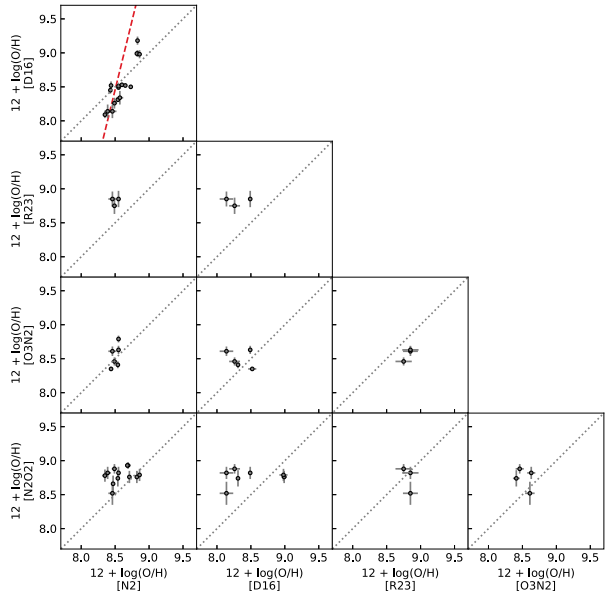

**Figure B1.** Scatter plots of integrated metallicities from the five available indicators, N2, D16, O3N2, R23, and N2O2. The dotted lines are $y = x$ lines. The dashed line in the first panel shows the best fit to the KURVS-CDFS galaxies only reported in S. Gillman et al. (2022) recalibrated to F. Bian et al. (2018) N2.

## APPENDIX C: BINNING SCHEME

In this appendix, we focus on discussing the effects on the choices of different binning schemes. In Section 4, we choose a binning scheme of three 0.5 arcsec radial bins. To investigate if the binning scheme brings significant influences on the results, we do a test of changing the bin width and number, to two 0.8 arcsec radial bins, and to the one proposed in S. Gillman et al. (2022). We note that once we choose two 0.8 arcsec radial bins there will not be the case of least-$\chi^2$ fitting. In the both schemes we reject the radial bins where any line flux required in the metallicity indicator has S/N < 4. Fig. C1 shows the comparison of the measured metallicity gradients using three 0.5 arcsec bins and two 0.8 arcsec bins, and Fig. C2 shows the comparison of the measured metallicity gradients using three 0.5 arcsec bins and the half-light radius binning scheme in S. Gillman et al. (2022). The Pearson correlation between metallicity gradients from three 0.5 arcsec bins and two 0.8 arcsec bins is $0.01 \pm 0.17$, while the Pearson correlation between metallicity gradients from three 0.5 arcsec bins and those of S. Gillman et al. (2022) is $0.10 \pm 0.24$. Thus, we adopt the scheme of three 0.5 arcsec bins.

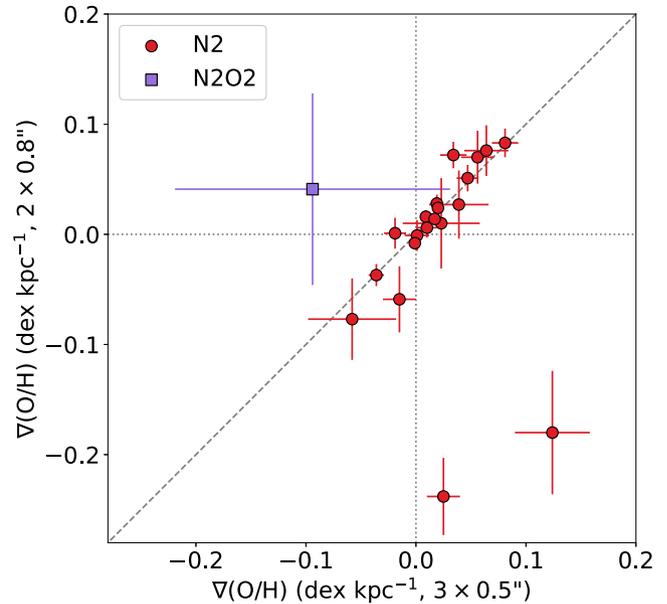

**Figure C1.** Comparison of the measured metallicity gradients using three 0.5 arcsec radial bins and two 0.8 arcsec radial bins setting a minimum S/N at 4. Circles and squares indicate metallicity gradients using N2 and N2O2, respectively. The dashed line, horizontal dotted line, and vertical dotted line illustrate $y = x$, $y = 0$, and $x = 0$, respectively. The Pearson correlation between the metallicity gradients using the two binning schemes is $0.01 \pm 0.17$.

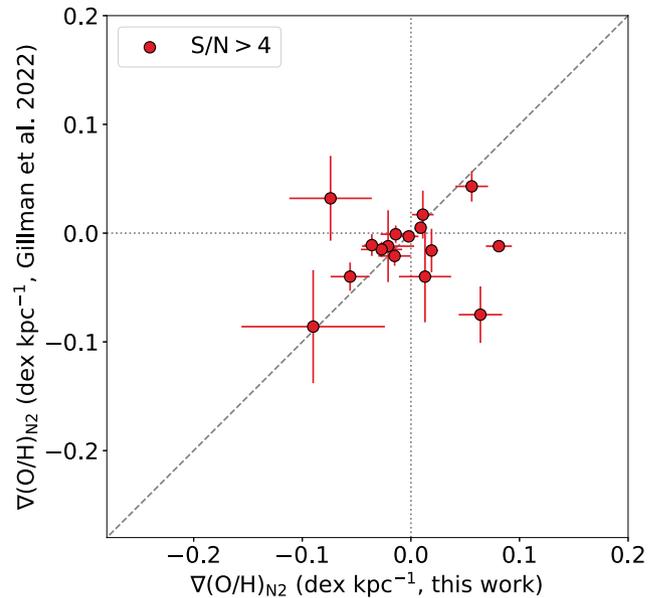

**Figure C2.** Comparison of the measured metallicity gradients using three 0.5 arcsec radial bins and those of S. Gillman et al. (2022), setting a minimum S/N of 4. To keep consistency both metallicities are estimated using N2. The dashed line, horizontal dotted line, and vertical dotted line illustrate $y = x$, $y = 0$, and $x = 0$, respectively. The Pearson correlation between the metallicity gradients using the two binning schemes is $0.10 \pm 0.24$.

This paper has been typeset from a T<sub>E</sub>X/L<sup>A</sup>T<sub>E</sub>X file prepared by the author.